\RequirePackage[2018-12-01]{latexrelease}

\documentclass{rQUF2e}

\usepackage{epstopdf}
\usepackage{subfigure}

\usepackage{enumerate}
\usepackage{multirow}
\usepackage{mathtools}
\usepackage{booktabs}
\usepackage{graphicx}
\usepackage{xcolor}
\usepackage{todonotes}
\usepackage{dsfont}
\usepackage[normalem]{ulem}

\theoremstyle{plain}
\newtheorem{theorem}{Theorem}[section]

\newtheorem{proposition}[theorem]{Proposition}

\theoremstyle{definition}

\theoremstyle{remark}

\begin{document}
	

\title{Risk-Neutral Generative Networks}

\author{Zhonghao Xian$^\star$$\dag$, Xing Yan$^\star$$\ddag$\thanks{$^\star$Co-first authors.}, Cheuk Hang Leung$\dag$, and Qi Wu$^{\ast}$$\dag$
\thanks{$^\ast$Corresponding author. Email: qiwu55@cityu.edu.hk}\\
\affil{$\dag$Department of Data Science, City University of Hong Kong\\
	$\ddag$Beijing Institute of Mathematical Sciences and Applications (BIMSA)} \received{v2.1 released October 2014} }

\maketitle

\begin{abstract}
We present a generative approach to price options and extract risk-neutral densities from the market. Specifically, we model the underlying log-returns on the time-to-maturity continuum as a generative model from standard normal. Neural nets are used to represent the term structures of the location, the scale, and the higher-order moments. We impose stringent conditions on the learning process to ensure no arbitrage. This model allows for the efficient generation of samples to price options across strikes and maturities. We have validated the effectiveness of this approach by benchmarking it against a comprehensive set of baseline models. Experiments show that the extracted risk-neutral densities accommodate a diverse range of shapes. Its accuracy significantly outperforms the extensive set of baseline models—including three parametric models and nine stochastic process models—in terms of accuracy and stability. The success of this approach is attributed to its capacity to offer flexible term structures for risk-neutral skewness and kurtosis.
\end{abstract}


\begin{classcode}G13\end{classcode}

\section{Introduction}\label{sec:Introduction}


The risk-neutral pricing theory was developed in the seventies when \cite{COX1976145} simplified the option pricing under the Black-Scholes model to an expectation calculation under a risk-neutral measure. This measure integrates the market consensus on future asset prices, adjusted for market risk preferences. The risk-neutral measure effectively adjusts the real-world probabilities of outcomes to reflect a world where all investors are risk-neutral. 
Since then, risk-neutral pricing models have become essential tools supporting the modern operation of the options market. Broker-dealers employ them to consistently hedge their option inventories in order to comply with stringent regulatory requirements on net risk exposures. Investors leverage them to identify arbitrage opportunities by calibrating a set of options across different strikes or maturities, assessing non-calibrated options, and exploring discrepancies for profitable investments. The fundamental task of constructing any risk-neutral pricing model ultimately boils down to the extraction of the risk-neutral density from market prices, as the risk-neutral density encapsulates the entirety of information inherent in the traded prices, both completely and exclusively \citep{cont1997beyond}. 

Researchers have explored the possibility of modelling the distribution for the underlying asset at maturity, or the implied volatilities among other quantities of direct interests, without resorting to any dynamics of the underlying. For this approach to work, no-arbitrage conditions need to be imposed during calibration to ensure the model is free from, at least, static arbitrage opportunities \citep{davis2007range, schweizer2008arbitrage, fengler2009arbitrage, follmer2011stochastic, cont2023simulation}. 
Prior methods taking this approach vary on their degrees of parameterization of the distribution. While parametric methods offer structural clarity and computational efficiency, non-parametric specifications are more adaptive to data, particularly when the true distribution has an irregular shape, contains multiple peaks, or has asymmetric heavy tails, but comes at the cost of losing interpretability and increasing computational complexity. However, arbitrage-free is more important than the level of parameterization because it safeguards the model against producing opportunities for riskless profit, which, in theory, should not exist in a complete and efficient market.  

What we are after is a generative approach that can balance the parameter interpretability with the data adaptability while rigorously imposing no-arbitrage conditions. In fact, prior works have explored the nonlinear capabilities of neural networks to approximate the option price or the volatility surface as a function of the underlying, the strike, and the maturity \citep{hutchinson1994nonparametric, garcia2000pricing, NIPS2000_44968aec, amilon2003neural, Yang_Zheng_Hospedales_2017, ackerer2020deep}. For our purpose, a straightforward thought is also using neural nets to represent the risk-neutral density, for example, following the treatment in the Generative Adversarial Network (GAN) \citep{goodfellow2014generative}, and letting the calibration process learn the generative network given the observations of option prices. However, neural networks, often viewed as ``black boxes'', typically lack interpretability.

To reinstall interpretability into the neural-net-based specification, we propose to model the random log-returns on the time-to-maturity continuum as a generative model. Specifically, we assume the underlying log-return is a deterministic function of two variables, the standard normal random variable and the time-to-maturity deterministic variable. In our specification, the model degenerates to a flexible random variable transforming the standard normal when the time-to-maturity is held constant.
This specification is structurally clear in that it allows flexible term structure shapes of the location parameter, the scale parameter, and the higher-order moments. It is both data adaptive and interpretable. Moreover, the generative nature allows one to generate samples efficiently to compute statistical quantities and to price options at any strikes and maturities. 

We call this specification the \textbf{R}isk-\textbf{N}eutral \textbf{G}enerative \textbf{N}etwork (RNGN), and showcase three specific representations: the risk-neutral quantile model (RN-Q) for single maturity calibration, the risk-neutral multi-layer perceptron model (RN-MLP) for multi-maturity calibration, and the risk-neutral double multi-layer perceptrons model (RN-DMLP) for multi-maturity calibration with additional shape complexities. As we make no assumptions of the price dynamics or the distributional families, we shall impose no-arbitrage conditions during calibration that are known in the literature, including three sets of inequality constraints on the partial derivatives, two sets of boundary conditions, and one set of pricing bounds. 

The empirical experiments show that the extracted density using the RNGN approach can accurately recover a wide range of shapes. It performs outstandingly well against an extensive set of baseline models, including three parametric models and nine stochastic process models. In particular, the RN-DMLP model generally achieves the lowest mean squared error (MSE) on both in-sample and out-of-sample datasets, even in scenarios involving options with extreme moneyness. Comprehensive testing further validates its superior stability. We attribute the success of RN-DMLP to its ability to offer flexible term structures for risk-neutral skewness and kurtosis.

The rest of the paper is organized as follows. Section \ref{sec:ProblemFormulation} formulates the problem of risk-neutral pricing and risk-neutral distribution extraction and revisits the no-arbitrage constraints. Section \ref{sec:OurMethods} introduces the generative model specification in our approach. Section \ref{sec:NumericalStudies} demonstrates the accuracy and stability of the proposed risk-neutral generative network through the simulation and empirical studies. Section \ref{sec:HigherOrderMoments} provides insights into the higher-order moments of the learned risk-neutral densities. Section \ref{sec:Conclusion} concludes the paper.

\section{No-Arbitrage Conditions}\label{sec:ProblemFormulation}


The time-$t$ prices of European calls and puts on an underlying $S_t$, struck at $K$ with time-to-maturity $\tau:=T-t$, are conditional expectations of the discounted payoff under the spot risk-neutral measure $\mathbb{Q}$:
\begin{equation}
	\begin{aligned} \label{eq:payoffF-St}
		C(S_t, K, \tau) & := \mathbb{E}^{\mathbb{Q}}[e^{-r\tau}(S_T-K)^+|S_t]  \\
		& = e^{-r\tau} \int_0^\infty (s - K)^+ f_{S_T}(s) ds, \\
		P(S_t, K, \tau) & := \mathbb{E}^{\mathbb{Q}}[e^{-r\tau}(K-S_T)^+|S_t]   \\
		& = e^{-r\tau} \int_0^\infty (K-s)^+ f_{S_T}(s) ds,
	\end{aligned}
\end{equation}
$\forall r>0,~ K> 0,~ T\geq t$. $f_{S_T}(s)$ is the risk-neutral density function of $S_T$ at maturity conditional on $S_t$, $(\cdot)^+$ denotes $\max\{\cdot, 0\}$. We assume the discount rate $r$ is a positive constant and there is no dividend payout. Alternatively, we can express \eqref{eq:payoffF-St} through the log-return $X_\tau \coloneqq X_{t:T} = \ln (S_T / S_t)$ and its associated density $q_{X_\tau}(x):=S_t e^x f_{S_T}(S_t e^x)$ as
\begin{equation}
	\begin{aligned}
		C(S_t, K, \tau) 
		&= e^{-r\tau}S_t \int_{-\infty}^{+\infty} \bigg(e^x - S_t^{-1}K \bigg)^{+} q_{X_\tau}(x) dx, \\
		P(S_t, K, \tau) 
		&= e^{-r\tau}S_t \int_{-\infty}^{+\infty} \bigg(S_t^{-1}K - e^x \bigg)^{+} q_{X_\tau}(x) dx.\label{eq:payoffF-St2}
	\end{aligned}
\end{equation}
Once we know $q_{X_\tau}(x)$, we have the density $f_{S_T}(s) = \frac{1}{s} \cdot q_{X_\tau}(\ln \frac{s}{S_t})$ of $S_T$ in case it is needed.

We aim to estimate $q_{X_\tau}(x)$ from a set of observed prices of calls $\{C_{i}\}_{i=1}^{N_C}$ and puts $\{P_{j}\}_{j=1}^{N_P}$ by minimizing 
\begin{equation}
	\frac{1}{N_C} \sum_{i=1}^{N_C} w^C_{i} L(C_i, \hat{C}_i) + \frac{1}{N_P} \sum_{j=1}^{N_P}  w^P_{j} L(P_j, \hat{P}_j),
\end{equation}
where $L(\cdot, \cdot)$ is the loss function, for example, the absolute or relative mean square error (MSE); $\hat{C}_i$ and $\hat{P}_i$ are the option prices given by the model; $w^C_{i}$ and $w^P_{i}$ are the regularized weights that may be related to the option liquidity levels and we set $w^C_{i}=w^P_{i}=1.0$ for simplicity.

In the calibration process, we shall impose the following six no-arbitrage conditions that are known to be both sufficient and (close to) necessary \cite[Theorem 2.1]{roper2010arbitrage} to see what conditions the learned $q_{X_\tau}(x)$ should satisfy:
\begin{enumerate}
	\item $\frac{\partial C}{\partial K} \leq 0$ and $\frac{\partial P}{\partial K} \geq 0$; \label{C1}
	\item $\frac{\partial^2 C}{\partial K^2} \geq 0$ and $\frac{\partial^2 P}{\partial K^2} \geq 0$; \label{C2}
	\item $\lim_{K \rightarrow \infty} C(S_t, K, \tau) = C(S_t, \infty,\tau) = 0$ and $\lim_{K \rightarrow 0} P(S_t, K, \tau) = P(S_t, 0,\tau) = 0$; \label{C3}
	\item $C(S_t, K, 0) = \left(S_t - K\right)^+$ and $P(S_t, K, 0) = \left(K - S_t\right)^+$ when $\tau=0$; \label{C4}
	\item $\frac{\partial C}{\partial \tau} \geq 0$ and $\frac{\partial P}{\partial \tau} \geq 0$; \label{C5}
	\item $\left(S_t - Ke^{-r\tau}\right)^+ \leq C(S_t, K, \tau) \leq S_t$ and $\left(Ke^{-r\tau} - S_t\right)^+ \leq P(S_t, K, \tau) \leq Ke^{-r\tau}$. \label{C6}
\end{enumerate}

\paragraph{Constraint 1}
The first condition requires the absence of call spread arbitrage among option strikes for any fixed option maturity. From Eqn. \eqref{eq:payoffF-St2}, we have
\begin{equation*}
	\begin{aligned}
		\frac{\partial C(S_{t}, K, \tau)}{\partial K} = -e^{-r\tau} \int_{\ln(K/S_t)}^{+\infty} q_{X_\tau}(x) dx, \quad \forall K> 0,~ \tau \geq 0.
	\end{aligned}
\end{equation*}
For $\frac{\partial C}{\partial K}\leq 0$ to be uniformly non-positive, the integral against $q_{X_\tau}(x)$ on the right-hand side needs to be uniformly non-negative, given the negative sign in front of the right-hand side and the fact that the discount factor $e^{-r\tau}$ is positive. Since $q_{X_\tau}(x)$ is a density function by construction in our setup where the random variable $X_\tau$ will be specified as an explicit function of the standard normal (details to be given in the next section), $q_{X_\tau}(x)$ is point-wise non-negative, implying the integral against it is indeed non-negative uniformly. The case for $\frac{\partial P}{\partial K}$ is similar. Therefore, $\frac{\partial C}{\partial K}\leq 0$ and $\frac{\partial P}{\partial K} \geq 0$ always hold in our setup.

\paragraph{Constraint 2}
The second condition is to ensure the pricing function free from butterfly spread arbitrage among option strikes for any fixed option maturity. Note that
\begin{equation*}
	\begin{aligned}
		\frac{\partial^2 C(S_{t},K,\tau)}{\partial K^2} & = \frac{\partial }{\partial K}\bigg\{ -e^{-r\tau} \int_{\ln(K/S_t)}^{+\infty} q_{X_\tau}(x) dx\bigg\} \\
		& = \frac{1}{K} e^{-r\tau} q_{X_\tau}\bigg(\ln\bigg(\frac{K}{S_{t}}\bigg)\bigg), \quad \forall K>0,~ \tau \geq 0.
	\end{aligned}
\end{equation*}
For the same reason that $q_{X_\tau}(\cdot)$ is constructed as a density function which is point-wise non-negative, $\frac{\partial^2 C(S_{t},K,\tau)}{\partial K^2}$ shall be uniformly non-negative. This is also the case for puts. Therefore, $\frac{\partial^2 C}{\partial K^2} \geq 0$ and $\frac{\partial^2 P}{\partial K^2} \geq 0$ always hold. 

\paragraph{Constraint 3}
The third condition imposes the boundary prices to be zero when the strike price of call options is set to infinity and the strike price of put options is set to zero. It is interpreted as that a put (or call) option has no value if its strike is zero (or infinitely high) because real-world stock prices cannot fall below zero or rise to infinitely high. 
More specifically, $\max(s-K,0)|_{K=+\infty} =0$ holds for the call case and $\max(K-s,0)|_{K=0} =0$ holds for the put case in Eqn. \eqref{eq:payoffF-St}, where $s$ is any possible realized value of $S_T$. Therefore, this constraint is always satisfied in our setup.


\paragraph{Constraint 4}
The fourth one sets price conditions on the zero time-to-maturity boundary. It is equivalent to requiring the random variable $X_\tau$ is degenerate, i.e., it almost surely takes a specific value, $X_0$, as the option's maturity is infinitely close to the current calendar time. In other words, its density $q_{X_\tau}(x)$ becomes a Dirac delta function,
\begin{equation*}
	\begin{aligned}
		& \lim_{\tau \downarrow 0} q_{X_\tau}(x) = \delta(x-X_0),~ \textrm{where} \
		X_0:=\lim_{t\leftarrow T}\ln (S_T/S_t) = 0.
	\end{aligned}
\end{equation*}
The Dirac delta function $\delta(x-X_0)$ is zero everywhere except at $x=X_0$. It has the property that the integral of the Dirac delta function against any continuous test function $\phi(x)$ picks out the value of the test function at $X_0$: $\int_{\mathbb{R}} \phi(x)\delta(x-X_0)dx = \phi(X_0)$. 
For $q_{X_\tau}(x)$ to be a Dirac delta function, it means $X_\tau$ takes on the value $X_0$ with probability one almost surely.
Therefore, option prices on the zero time-to-maturity boundary are equal to their payoffs.
In the next section, we shall show that our specifications for $X_\tau$ have made sure that $X_\tau$ indeed takes on the value $X_0$ with probability one as $\tau$ approaches the zero boundary.
Therefore, this constraint is satisfied in our setup.

\paragraph{Constraint 5}
The fifth condition is equivalent to the absence of calendar spread arbitrage between adjacent maturities for any fixed strike. It demands that the pricing functions, i.e. \eqref{eq:payoffF-St2}, should be monotonically increasing as the time-to-maturity extends, upon fixing the rest of other contractual and model parameters. It reflects the economic principle that an optionality should be more valuable if it gives the holder a higher chance to be potentially more profitable. In the case of a call (or put) option, a longer maturity implies there is more time, hence a higher probability, for the underlying stock to surpass (or fall below) a given threshold. We shall impose this condition as a soft constraint through a regularization term when calibrating the model, as detailed in Section \ref{sec:OurMethods}. 


\paragraph{Constraint 6}

The sixth condition imposes upper and lower bounds on the prices of European options.  Together with the non-negativity of option prices, these pricing bounds are direct consequences of assuming the following equality (holds uniformly for any time-to-maturity):
\begin{equation}
	\ln \mathbb{E}^{\mathbb{Q}}[ e^{X_\tau}]  =r\tau, \quad \forall \tau >0.
\end{equation}
It is equivalent to requiring that the present value of the stock price at any future time, as a conditional expectation under the spot risk-neutral measure, equals the current stock price,  
\begin{equation}
	\mathbb{E}_t^Q[S_T] = S_t e^{r(T-t)}, \quad \forall T>t.
\end{equation}

The argument is as follows. First, if $\mathbb{E}_t^Q[S_T] = S_t e^{r(T-t)}$ holds for all $T>t$, then the put-call parity holds,
\begin{equation*}
	\begin{aligned}
		&\qquad C(S_t,K,\tau) - P(S_t,K,\tau) \\
		& =e^{-r\tau} \int_0^{+\infty} \big[ (s-K)^+ -(K-s)^+\big] f_{S_T}(s) ds \\
		& = e^{-r\tau} \int_0^{+\infty} (s-K) f_{S_T}(s) ds \\
		& = e^{-r\tau}\mathbb{E}_t^Q[S_T] - Ke^{-r\tau} =S_t - Ke^{-r\tau}, \quad \forall \tau >0,~ K>0.
	\end{aligned}
\end{equation*}

The lower bound of calls is obtained by applying the put-call parity and using the fact that option prices are non-negative,
\begin{equation*}
	\begin{aligned}
		C(S_t,K,\tau) = P(S_t,K,\tau) + S_t - Ke^{-r\tau}\geq S_t - Ke^{-r\tau}.
	\end{aligned}
\end{equation*}
Together with $C(S_t,K,\tau)\geq 0$, we have calls bounded below as $C(S_t, K, \tau)\ge (S_t - Ke^{-r\tau})^+, ~\forall \tau>0,~ K>0$.
The upper bound of calls is the consequence of putting Contraint 1, i.e. $\frac{\partial C}{\partial K} \leq 0$, together with the equality we assume to hold, i.e. $\mathbb{E}_t^Q[S_T] = S_t e^{r(T-t)}, ~\forall T>t$,
\begin{equation*}
	\begin{aligned}
		& C(S_t, K, \tau)\le C(S_t, 0, \tau),\quad \textrm{where} \\
		& C(S_t,0,\tau) = e^{-r\tau} \int_0^{+\infty} s f_{S_T}(s) ds = e^{-r\tau} \mathbb{E}_t^Q[S_T] = S_t.
	\end{aligned}
\end{equation*}
Therefore, calls are bounded above as $C(S_t, K, \tau)\le C(S_t, 0, \tau)= S_t,~ \forall K>0, ~\tau>0$. 

%
%

In the case of put options, the argument is similar. First, we already know from the upper bound of calls that $C(S_t,K,\tau)-S_t\le 0$. Therefore, the puts are bounded above by $Ke^{-r\tau}$ from the put-call parity because $P(S_t,K,\tau)=Ke^{-r\tau}+C(S_t,K,\tau) -S_t\le Ke^{-r\tau}$. Meanwhile, by the non-negativity of option prices, both $P(S_t,K,\tau)\geq 0$ and $P(S_t,K,\tau)=Ke^{-r\tau}+C(S_t,K,\tau) -S_t \geq Ke^{-r\tau}-S_t$ should hold. Put them together and we have the puts bounded below by $(Ke^{-r\tau}-S_t)^+$. 

Rechecking the equality $\mathbb{E}_t^Q[S_T] = S_t e^{r(T-t)}$, or equivalently, $C(S_t, 0, \tau) = S_t$, we argue that it should hold in a rational market because a call option with a zero strike will always be exercised, resulting in a payoff $S_T$. This is equivalent to possessing the underlying asset directly.

\textbf{To summarize}, Constraints \ref{C1}--\ref{C4} collectively require that the log-return random variable $X_\tau$ has a density degenerating to the Dirac delta function on the zero time-to-maturity boundary. Constraint \ref{C5} further regulates the density behavior of $X_\tau$ along the time-to-maturity dimension to ensure that options prices are non-decreasing functions of time-to-maturity. Constraint \ref{C6} requires that the expected future stock price at any future time should equal the current stock price grown at the risk-free rate. 


\section{Model Specification}\label{sec:OurMethods}

%

At different option maturity times, the log-returns $X_\tau = \ln (S_T / S_t)$ are scalar random variables, characterized by their individual marginal distributions. 
Our specification begins by assuming the log-return $X_\tau = \ln (S_T / S_t)$ on the time-to-maturity continuum is a function of two variables, the standard normal random variable $Z$ and the time-to-maturity deterministic variable $\tau\in \mathbb{R}^+$, 
\begin{equation}\label{eq_GeneralRND_complex}
	\begin{aligned}
		X_{\tau} &:= X(Z,\tau) = \mu(\tau) + \sigma(\tau) \cdot Z \cdot G(Z, \tau),\quad Z\sim \mathbb{N}(0,1),\\
		X_{0}&:=\lim_{t \leftarrow T}\ln(S_T/S_t)=0,
	\end{aligned}
\end{equation} 
where $\mu(\cdot),~\sigma(\cdot), ~G(\cdot,\cdot)$ are deterministic functions, either in closed forms or parameterized by neural nets with the Softplus activation function $\ln\left(1+\exp(x)\right)$. 

This specification is functionally flexible. Holding $\tau$ constant, $X(Z,\tau)$ degenerates to a flexible random scalar, representing the log-return at a specific time-to-maturity, with the randomness from standard normal $Z$\footnote{The scalar random variable $Z$ can be generalized to a stochastic (standard normal) curve $Z_\tau$ to highlight that different time-to-maturities will have different random drivers $Z_\tau$. This generalization can be used to model the dependence among returns across maturities.}. 
In the meantime, this specification remains structurally interpretable. In particular, $\mu(\tau)$ models the location term structure of the log-returns analogous to the mean, $\sigma(\tau)$ represents the scale term structure analogous to the volatility, and through $G(\cdot, \tau)$ the term structures of higher-order moments such as skewness and kurtosis. 

One can simply set $G(\cdot, \cdot)\equiv 1$, yielding normal distributions for $X_\tau \sim \mathbb{N}(\mu(\tau), \sigma^2(\tau))$, or as flexible as represented by neural nets to capture any non-Gaussian distributional properties of the financial data \citep{cont2001empirical}, such as having irregular shapes, containing multiple peaks, or exhibiting asymmetric heavy tails. The three term structure maps, $\tau \mapsto \mu(\tau)$, $\tau \mapsto \sigma(\tau)$, and $\tau \mapsto G(\cdot, \tau)$, can either be in closed forms for simplicity, or be parameterized by neural nets to cater to the real-world complexities of variations in the data distribution. 

In addition, this specification is generative by construction in that the functional form of $X(Z,\tau)$ is an explicit and deterministic function of the standard normal. For calibration, sample averaging shall approximate well the option prices,  
\begin{equation}\label{eqn:pricing_strategy}
	\begin{aligned}
		& \hat{C}(S_t, K, \tau)= e^{-r\tau}S_t \cdot\frac{1}{N}\sum_{n=1}^N \left(e^{X(Z_n,\tau)} - S_t^{-1}K \right)^{+}, \\
		& \hat{P}(S_t, K, \tau) = e^{-r\tau}S_t \cdot\frac{1}{N}\sum_{n=1}^N \left(S_t^{-1}K - e^{X(Z_n,\tau)} \right)^{+}.
	\end{aligned}
\end{equation}
Here $Z_n$ is a sample of $Z$, and thereby $X(Z_n,\tau)$ is a sample of the log-return $X_{\tau}$. Once the model parameters are learned, computing statistics of interest is also straightforward. The only simulation needed is a set of standard normal samples. 

Finally, there exists a continuous density for $X(Z,\tau)$ specified by Eqn. \eqref{eq_GeneralRND_complex}, when $G(\cdot,\cdot)$ in Eqn. \eqref{eq_GeneralRND_complex} is represented by multi-layer perceptrons (MLPs) with the Softplus activation. If so, $G(\cdot,\cdot)$ consists of a finite number of simple mathematical operations, then $X(\cdot,\tau)$ is a piecewise (strictly) monotonic function when fixing $\tau$. Applying the general form of the change-of-variable formula \cite[Theorem 2.1.8]{casella2024statistical}, we can conclude that the random variable $X(Z,\tau)$ has a continuous density.

In the upcoming subsections, we showcase three different specifications for $X_{\tau}$, one for single maturity calibration, one for multi-maturity calibration, and one for multi-maturity calibration with additional shape complexities. For these three particular specifications, Constraints \ref{C1}--\ref{C4} hold by construction; Constraint \ref{C5} shall be enforced as inequality constraints and Constraint \ref{C6} as an equality constraint requiring
\begin{equation}\label{eq:noarbitrage-general}
	\ln \mathbb{E}^{\mathbb{Q}}\Big[ \exp\Big(\mu(\tau) + \sigma(\tau) \cdot Z \cdot G(Z, \tau)\Big)\Big]  =r\tau, \quad \forall \tau >0.
\end{equation}

\subsection{Single Maturity}\label{par:GML-Q}

When there is only one maturity, there is no notion of term structure. Now $X_\tau$ becomes a random scalar, and we can drop the subscript $\tau$ in $X_{\tau}$ and set 
\begin{equation}
	\mu(\tau) := \mu,~~\sigma(\tau) :=\sigma,~~ G(Z,\tau) := G(Z).
\end{equation}
If we assume $G(Z)= \frac{u^{Z}}{A} + \frac{v^{-Z}}{A} + 1$, we have the log-return $X$ taking the following form, 
\begin{equation}\label{eqt:ref_quantile}
	X = \mu + \sigma Z\left(\frac{u^{Z}}{A} + \frac{v^{-Z}}{A} + 1\right),
\end{equation}
where $\mu$ and $\sigma$ are the location and scale parameters; $u\ge 1$ controls the right tail and $v\ge 1$ controls the left tail; $A$ is a positive constant. Setting $u=v=1$ turns $X$ into a normal distribution as $G(Z)$ becomes a constant. This specification recovers the model introduced in \cite{yan2019cross}, which was proposed to capture the asymmetric heavy tail nature of asset returns, motivated by the parsimonious heavy-tailed quantile function ideas \citep{NEURIPS2018_9e3cfc48}.
In Appendix \ref{appen:htqf}, we provide some statistical properties of the distribution associated with the random variable in Eqn. \eqref{eqt:ref_quantile}, with a particular focus on its moment structure. 

Given the model parameters $\left\{\mu, \sigma, u, v\right\}$, the market observables $\left\{S_t, r\right\}$, and the contractual parameters $\left\{K^C,\tau\right\}$ or $\left\{K^P,\tau\right\}$ for the call or the put, we obtain the model prices by drawing $N$ standard normal samples $\{Z_n\}$ and averaging the discounted payoffs,
\begin{equation}
	\begin{aligned} \label{eq:pricing_formula-Q}
		&\hat{C}_{Q} = \frac{1}{N}\sum_{n=1}^N e^{-r\tau} S_t \left( e^{X(Z_n)} - S_t^{-1}K^C \right)^+,\\
		&\hat{P}_{Q} = \frac{1}{N}\sum_{n=1}^N e^{-r\tau} S_t \left( S_t^{-1}K^P - e^{X(Z_n)} \right)^+,
	\end{aligned} 
\end{equation}
where
\begin{equation}
	X(Z_n) =\mu+\sigma Z_n\left(\frac{u^{Z_n}}{A}+\frac{v^{-Z_n}}{A}+1\right). \label{eq:log-return-Q} 
\end{equation}

The log-return as specified in Eqn. \eqref{eqt:ref_quantile} always attains a density, therefore Constraints \ref{C1}-\ref{C3} shall hold. Constraints \ref{C4} and \ref{C5} are not applicable for the single maturity case as they concern price behaviors with changing time-to-maturity. For Constraint \ref{C6}, we drive from Eqn. \eqref{eq:noarbitrage-general} that the model parameters $\left\{\mu, \sigma, u, v\right\}$ should satisfy the following equality as a hard constraint at a fixed $\tau$,
\begin{equation}\label{eq:noarbitrage-Q}
	\mu + \ln\left[\frac{1}{N}\sum^N_{n=1}\exp\Big(\sigma Z_n \big(\frac{u^{Z_n}}{A}+\frac{v^{-Z_n}}{A}+1\big)\Big)\right] = r\tau, 
\end{equation}
which is equivalent to Eqn. (\ref{eq:noarbitrage-general}) after replacing the expecation by the  sample average.
Therefore, one only needs to calibrate the three parameters $\sigma, u, v$. 

We call this specification, Eqns. \eqref{eq:pricing_formula-Q}--\eqref{eq:noarbitrage-Q}, the RN-Q (\textbf{R}isk-\textbf{N}eutral \textbf{Q}uantile Model) and shall use it as the base model for later studies. 

\subsection{Multiple Maturities}


When there are multiple maturities to calibrate, we shall set the functions $\mu(\tau)$, $\sigma(\tau)$, and $G(Z,\tau)$ as,
\begin{equation}
	\begin{aligned}
		&\mu(\tau) := r \cdot\tau \cdot G^\mu(\tau;\theta^\mu),\\
		&\sigma(\tau) :=\sigma\sqrt{\tau},\\
		&G(Z,\tau):=G^Z(Z;\theta^Z) + G^\tau(\tau;\theta^\tau) +1,
	\end{aligned}
\end{equation}
where $ G^\mu,~ G^Z,~ G^\tau$ are MLP neural nets, parameterized by $\theta^\mu,~\theta^Z,~\theta^\tau$; $Z$ and $\tau$ are the same standard normal variable and the time-to-maturity variable as before. The resulting log-return variable is as follows,
\begin{equation}\label{eq:function-X-MLP}
	\begin{aligned}
		X_\tau & :=X(Z, \tau;\theta^{M}) = r \tau G^\mu(\tau;\theta^\mu) \\
		& + \sigma\sqrt{\tau}  Z   [G^Z(Z;\theta^Z) + G^\tau(\tau;\theta^\tau) + 1], \quad \forall \tau>0,\\
		X_0 & := X(Z, \tau;\theta^{M})|_{\tau=0} = 0.
	\end{aligned}
\end{equation}
In our model, all hidden layers in the subnetworks $G^\mu(\tau;\theta^\mu)$, $G^Z(Z;\theta^Z)$, and $G^\tau(\tau;\theta^\tau)$ consist of a linear layer followed by the Softplus activation. For the final layers, we adopt different activation choices depending on the modeling requirements: 
\begin{itemize}
	\item The final layer of $G^\mu(\tau;\theta^\mu)$ does not use Softplus, allowing the drift-related term to take both positive and negative values;
	\item The final layers of $G^Z(Z;\theta^Z)$ and $G^\tau(\tau;\theta^\tau)$ use Softplus to ensure positive outputs in these components.
\end{itemize}

Note that the above specifications are by no means the only choice. 
In our specification, we impose an additive structure in the function $G(Z, \tau)$, where the first component  $G^Z(Z;\theta^Z)$ depends only on $Z$ as the input, and $G^\tau(\tau;\theta^\tau)$ depends only on $\tau$ as the input. This structure facilitates the derivation of the first-order partial derivatives with respect to the variables $Z$ and $\tau$ when learning the parameters of the neural nets. 

We shall term this specification the RN-MLP model (\textbf{R}isk-\textbf{N}eutral Generative Network with \textbf{M}ulti-\textbf{L}ayer \textbf{P}erceptron), under which the pricing formulas are as follows, $\forall \tau>0$,
\begin{equation}
	\begin{aligned} \label{eq:pricing_formula-MLP}
		\hat{C}_{M} &= S_t e^{-r\tau} \frac{1}{N}\sum_{n=1}^{N}\bigg(e^{X(Z_n, \tau;\theta^{M})} - S_t^{-1}K^C \bigg)^+,\\ 
		\hat{P}_{M} &= S_t e^{-r\tau} \frac{1}{N}\sum_{n=1}^{N}\bigg(S_t^{-1}K^P - e^{X(Z_n, \tau;\theta^{M})} \bigg)^+, 
	\end{aligned}
\end{equation}
where $Z_n$ is the $n$-th sample of $Z$ and $\theta^{M} = \{\theta^\mu, \theta^Z, \theta^\tau, \sigma\}$ is the trainable parameter set. 

From the discussions on the density before, we know \eqref{eq:function-X-MLP} attains a density, therefore Constraints \ref{C1}--\ref{C3} hold. We let the location term $\mu(\tau) = r\tau G^\mu(\tau;\theta^\mu)$ and the scale term $\sigma(\tau)=\sigma\sqrt{\tau}$ depend on $\tau$ explicitly to ensure $X_\tau$ does indeed degenerate to $0$ at $\tau=0$ to satisfy Constraint \ref{C4}. 
In Proposition \ref{prop:GML-MLP}, we introduce conditions under which the model prices $\hat{C}_{M}$ and $\hat{P}_{M}$, given by Eqn. \eqref{eq:pricing_formula-MLP}, satisfy Constraints \ref{C5} and \ref{C6}, therefore are free from arbitrage. The proof can be found in Appendix \ref{sec:appendix1}.
\begin{proposition}\label{prop:GML-MLP}
	Let the log-return $X_\tau$ follow the specification given by Eqn. \eqref{eq:function-X-MLP}. The associated option prices $\hat{C}_{M}$ and $\hat{P}_{M}$ in Eqn. \eqref{eq:pricing_formula-MLP} are free from arbitrage if and only if 1) $\tilde{J}^\tau_{C}(\tau,K^{C},\theta^{M})\geq 0$, 2) $\tilde{J}^\tau_{P}(\tau,K^{P},\theta^{M})\geq 0$, and 3) $J^\mu_{M}(\tau,\theta^{M})=0$ where $\theta^{M} = \{\theta^\mu, \theta^Z, \theta^\tau, \sigma\}$ and
	\begin{equation} \label{eqt:J-MLP}
		\begin{aligned}
			\tilde{J}^\tau_{C}(\tau,K^{C},\theta^{M}) & \coloneqq 
			\frac{1}{N} \sum_{n=1}^{N} \mathds{1}_{\{e^{X(Z_n, \tau;\theta^{M})} \geq \frac{K^C}{S_t}\}}\times\\
			&\left[\left({\partial_\tau X(Z_n, \tau;\theta^{M})} - r\right) e^{X(Z_n, \tau;\theta^{M})} + r \frac{K^C}{S_t}\right],  \\    
			\tilde{J}^\tau_{P}(\tau,K^{P},\theta^{M}) &\coloneqq 
			\frac{1}{N} \sum_{n=1}^{N} \mathds{1}_{\{e^{X(Z_n, \tau;\theta^{M})} \leq \frac{K^P}{S_t}\}}\times \\
			& \left[\left(r - {\partial_\tau X(Z_n, \tau;\theta^{M})}\right) e^{X(Z_n, \tau;\theta^{M})} - r \frac{K^P}{S_t}\right],\\
			J^\mu_{M}(\tau,\theta^{M}) &\coloneqq 
			\bigg| r\tau \left( G^\mu(\tau;\theta^\mu) -1 \right) \\
			& + \ln\left(\frac{1}{N} \sum_{n=1}^N e^{\sigma\sqrt{\tau} \cdot Z_n \cdot [G^Z(Z_n;\theta^Z) + G^\tau(\tau;\theta^\tau) + 1]}\right) \bigg|^2. 
		\end{aligned}
	\end{equation}
\end{proposition}

Consequently, to train the RN-MLP model, we optimize the following objective:
\begin{equation}\label{eq:objective function-MLP_constrained} 
	\begin{aligned}
		\underset{\theta^{M}}{\text{min}}\quad
		& \mathcal{L}_{M} := \frac{1}{N_C} \sum_{i=1}^{N_C}L(C_i, \hat{C}_i) + \frac{1}{N_P} \sum_{{j}=1}^{N_P} L(P_{j}, \hat{P}_{j}),  \\
		\text{s.t. }
		&  ~\tilde{J}^{\tau}_{C}(\tau,K^{C},\theta^{M}) \geq 0,~
		\tilde{J}^{\tau}_{P}(\tau,K^{P},\theta^{M}) \geq 0,~
		J^\mu_{M}(\tau,\theta^{M})  = 0.
	\end{aligned}
\end{equation}
$L$ is the loss function, for example, either the absolute or the relative mean square error (MSE); $\{C_i,P_j\}$ are the observed market prices of options; and $\{\hat{C}_i,\hat{P}_j\}$ are the model prices. In practice, the above constrained optimization problem is rather difficult to solve with multi-layer neural networks. We can reformulate the problem \eqref{eq:objective function-MLP_constrained} as an unconstrained one through penalized relaxation,
\begin{equation}\label{eq:objective function-MLP}
	\begin{aligned}
		\underset{\theta^{M}}{\text{min}}\quad
		\mathcal{L}_{M} & :=  \frac{1}{N_C} \sum_{i=1}^{N_C} L(C_i, \hat{C}_i) + \frac{1}{N_P} \sum_{{j}=1}^{N_P} L(P_{j}, \hat{P}_{j}) + \lambda\cdot J(\theta^{M}), \\
		J(\theta^{M}) & := \sum_{\{\tau,K^{C}\}}\left(-\tilde{J}^{\tau}_{C}(\tau,K^{C},\theta^{M})\right)^{+} \\
		&+\sum_{\{\tau,K^{P}\}}\left(-\tilde{J}^{\tau}_{P}(\tau,K^{P},\theta^{M})\right)^{+} 
		+ \sum_{\{\tau\}} J_{M}^{\mu}(\tau,\theta^{M}). 
	\end{aligned}
\end{equation}
$J(\theta^{M})$ is the penalty term and $\lambda$ is the regularization hyper-parameter which can be set to one in the experiments.

\subsection{Additional Shape Flexibilities}
The RN-MLP model can flexibly model the log-returns with complex risk-neutral densities. However, sometimes the log-return exhibits a bi-modal or multi-modal distribution (e.g., see \cite{orosi2015estimating} and \cite{schmitt2017bimodality}).
Moreover, option-implied distributions exhibit rich complex characteristics, including abnormal tail behavior and sharpness/flatness in the central mass, especially during volatile market conditions.
Based on these considerations, we introduce two sources of randomness and enhance the model's flexibility through transformations of two Gaussian variables.
Now, we showcase how to extend the single RN-MLP specification into a mixture of two RN-MLPs to allow additional shape flexibilities,
\begin{equation} \label{eq:function-X-DMLP}
	\begin{aligned} 
		&X_\tau:=X(Z, \tau;\theta^{dM}) 
		= \alpha \cdot X_1(Z, \tau;\theta^{M}_1) + \\
		& \qquad\qquad (1 - \alpha) \cdot X_2(Z,\tau;\theta^{M}_2), \quad \forall \tau>0,\\
		& X_0 := X(Z, \tau;\theta^{dM})|_{\tau=0} = 0,
	\end{aligned}
\end{equation}
where $\alpha \in \mathbb{R}$, which is not limited in $[0, 1]$, and $X_i,~i=1,2$ share the same structure as that of RN-MLP in Eqn. \eqref{eq:function-X-MLP}, 
\begin{equation} \label{eq:function-X-DMLP2}
	\begin{aligned}
		&X_i(Z,\tau;\theta_i^M) = r\tau G_i^\mu(\tau;\theta_i^\mu)+\sigma_i  Q_{i}(Z,\tau;\theta^\sigma_i),\\
		\text{with } & Q_{i}(Z,\tau;\theta^\sigma_i)
		=\sqrt{\tau} Z  \left[G^Z_i(Z;\theta^Z_i) + G^\tau_i(\tau;\theta^\tau_i) + 1\right].
	\end{aligned}
\end{equation}
The set of trainable parameters is $\theta^{dM} = \{\alpha,\theta^{M}_1, \theta^{M}_2\}$ where $\theta^{M}_i= \{ \sigma_i, \theta^\mu_i, \theta_i^\sigma\}$ and $\theta_i^\sigma = \{  \theta^Z_i,\theta^\tau_i\}$ for $i=1,2$.

We call this specification the RN-DMLP model (\textbf{R}isk-\textbf{N}eutral Generative Network with \textbf{D}ouble \textbf{M}ulti-\textbf{L}ayer \textbf{P}erceptrons). The sample averaging formula of the option prices and the conditions for the model to be free of arbitrage are similar to those of the single RN-MLP case. $\forall \tau>0$,
\begin{equation}
	\begin{aligned}
		\begin{split} \label{eq:pricing_formula-DMLP}
			\hat{C}_{dM} = S_t e^{-r\tau} \frac{1}{N}\sum_{n=1}^{N}\left(e^{X(Z_n, \tau;\theta^{dM})} - S_t^{-1}K^C \right)^+, \\
			\hat{P}_{dM} = S_t e^{-r\tau} \frac{1}{N}\sum_{n=1}^{N}\left(S_t^{-1}K^P - e^{X(Z_n, \tau;\theta^{dM})} \right)^+.
		\end{split}
	\end{aligned}
\end{equation}

Similar to Proposition \ref{prop:GML-MLP}, we present in Proposition \ref{prop:GML-DMLP} the conditions under which $\hat{C}_{dM}$ and $\hat{P}_{dM}$ given in Eqn. \eqref{eq:pricing_formula-DMLP} satisfy Constraints \ref{C5} and \ref{C6}. The proof is analogous to that of Proposition \ref{prop:GML-MLP}, thereby is omitted.
\begin{proposition}\label{prop:GML-DMLP}
	Let $X_\tau$ follow the specification given by Eqn. (\ref{eq:function-X-DMLP}--\ref{eq:function-X-DMLP2}). The associated option prices $\hat{C}_{dM}$ and $\hat{P}_{dM}$ in Eqn. \eqref{eq:pricing_formula-DMLP} are free from arbitrage if and only if 1) $\tilde{J}^\tau_{C}(\tau,K^{C},\theta^{dM})\geq 0$, 2) $\tilde{J}^\tau_{P}(\tau,K^{P},\theta^{dM})\geq 0$, and 3) $J_{dM}^{\mu}(\tau, \theta^{dM}) =0$, where $\theta^{dM} = \{\alpha_i,\sigma_i,\theta^{\mu}_i, \theta^{\sigma}_i\}_{i=1}^{2}$, $\alpha_1+\alpha_2=1$, and 
	\begin{equation}\label{eqt:J-DMLP}
		\begin{aligned}
			&\tilde{J}^\tau_{C}(\tau,K^{C},\theta^{dM}) \coloneqq  
			\frac{1}{N} \sum_{n=1}^{N} \mathds{1}_{\{e^{X(Z_n, \tau;\theta^{dM})} \geq \frac{K^C}{S_t}\}}\times \\
			&\qquad \left[\left({\partial_\tau X(Z_n, \tau;\theta^{dM})} - r\right) e^{X(Z_n, \tau;\theta^{dM})} + r \frac{K^C}{S_t}\right],\\
			&\tilde{J}^\tau_{P}(\tau,K^{P},\theta^{dM}) \coloneqq 
			\frac{1}{N} \sum_{n=1}^{N} \mathds{1}_{\{e^{X(Z_n, \tau;\theta^{dM})} \leq \frac{K^P}{S_t}\}}\times \\
			&\qquad \left[\left(r - {\partial_\tau X(Z_n, \tau;\theta^{dM})}\right) e^{X(Z_n, \tau;\theta^{dM})}- r \frac{K^P}{S_t}\right],\\
			&J_{dM}^{\mu}(\tau, \theta^{dM}) \coloneqq 
			\bigg| r\tau\left( \sum_{i=1}^{2}\alpha_i G_i^\mu(\tau;\theta^\mu_i) -1 \right)
			+ \\
			&\qquad \ln\left(\frac{1}{N} \sum_{n=1}^N \exp\left[  \sum_{i=1}^{2} \alpha_i\sigma_i Q_{i}(Z_n, \tau;\theta^{\sigma}_i) \right] \right)\bigg|^2.	
		\end{aligned}
	\end{equation}
\end{proposition}

Using the same penalized relaxation technique we used to reformulate the problem \eqref{eq:objective function-MLP_constrained}, we can calibrate the RN-DMLP model as follows,
\begin{equation}\label{eq:objective function-DMLP}
	\begin{aligned}
		\underset{\theta^{dM}}{\text{min}}\quad
		&  \frac{1}{N_C} \sum_{i=1}^{N_C} L(C_i, \hat{C}_i) + \frac{1}{N_P} \sum_{{j}=1}^{N_P} L(P_{j}, \hat{P}_{j}) + \lambda \cdot J(\theta^{dM}),\\
		J(\theta^{dM}) &  :=\sum_{\{\tau,K^{C}\}}\left(-\tilde{J}^{\tau}_{C}(\tau,K^{C},\theta^{dM})\right)^{+}\\
		& + \sum_{\{\tau,K^{P}\}}\left(-\tilde{J}^{\tau}_{P}(\tau,K^{P},\theta^{dM})\right)^{+} + \sum_{\{\tau\}} J_{dM}^{\mu}(\tau,\theta^{dM}).
	\end{aligned}
\end{equation}

\subsection{Divided Payment}

Incorporating the dividend rate enhances the applicability of our method, extending its scope. Accordingly, we have revised our approach as follows. The modification to include the dividend rate is straightforward. In the absence of dividends, the variable of interest in the modeling throughout the paper is the log return, $X_\tau = \ln (S_T/S_t )$. When dividends are paid from holding the underlying asset, we adjust the definition of $X_\tau$ to:
$$
X_\tau = \text{the actual return from holding the underlying asset from time } t \text{ to } T.
$$
Let the dividend rate be denoted by $q_\tau$. We then have
\begin{equation}
	\ln (S_T/S_t) = X_\tau - q_\tau \tau.
\end{equation}
Under such circumstances, the option payoff, such as $(S_T - K)^+$, can be expressed as $S_t \left( e^{ X_\tau - q_\tau \tau} - \frac{K}{S_t} \right)^+$. Consequently, the pricing formula becomes:
\begin{equation}
	\mathbb{E}^\mathbb{Q} \left[ e^{-r\tau} S_t \left( e^{ X_\tau - q_\tau \tau} - \frac{K}{S_t} \right)^+  \bigg| S_t \right].
\end{equation}
The only distinction from the no-dividend case is the subtraction of $q_\tau \tau$ from $X_\tau$ in the pricing formula. This minor adjustment is consistent across all pricing formulas presented in the paper. This is the sole modification required to account for dividend payments.

\section{Experiments}\label{sec:NumericalStudies}

In this section, we undertake a comprehensive analysis comprising a simulation study and three empirical analyses to showcase the superior performance of our proposed models in terms of reliability, accuracy, stability, and alignment with financial intuitions.
Firstly, in the simulation study, we generate simulated option prices under the Heston assumption. Our results demonstrate that the three proposed models, namely RN-Q, RN-MLP, and RN-DMLP, are capable of accurately recovering the true RNDs across varying levels of skewness. This finding underscores the reliability of our models in extracting the RND.

Moving to the empirical studies, we conduct three experiments using S\&P 500 option prices over the extensive sample period from January 4, 1996, to February 28, 2023.
The first experiment evaluates the out-of-sample performance of our models. At each observation date, we first calibrate the proposed models using options with specific maturities and strikes. Subsequently, we employ the calibrated models to price the remaining options, such as those with extreme moneyness, to compare with observed option prices. We compare our models against twelve classical models, serving as baselines, to assess option pricing performance.
The second experiment investigates the stability of the recovered RNDs across all competing models, providing valuable insights into the robustness of our proposed models.
Lastly, the third experiment reveals the alignment of our models with financial insights.

\subsection{Hyperparameter Settings}
\label{par:hyperparameter} 

We outline the hyperparameters utilized in training the proposed models. While tuning or selecting hyperparameters can significantly enhance deep learning model performance, we aim to avoid dedicated tuning procedures and instead employ a unified hyperparameter setting, such as a unified network structure.
In the training of RN-Q model, we use $\sigma=0.2$, $u=1.1$, $v=1.1$ as initial values, and set $A=4.0$ as in the original paper. For RN-MLP and RN-DMLP models, the MLP architecture consists of two hidden layers, each comprising 32 neurons. The learning rate is fixed at $0.01$. The penalty hyper-parameter $\lambda$ is set to one.
Throughout the article, we maintain these configurations as the default setting without further fine-tuning, unless explicitly specified otherwise.

\subsection{Simulation Study}

In this simulation study, our objective is to assess the reliability of our proposed models in recovering the true RND from option data. We specify the contractual parameters as follows: the underlying asset price $S_t=1000$, the strike price $K^C$ ranging from $400$ to $1600$ in increments of $20$, a time-to-maturity $\tau=0.25$, and a risk-free rate $r=0.04$. This setup results in a total of $61$ options. To simulate option prices, we employ the Heston model \citep{heston1993closed}, a well-established stochastic volatility model extensively used in practice in the financial industry. The Heston model introduces a stochastic process of volatility to differ from the constant volatility assumption in the Black-Scholes model (BSM) and align with the observed volatility smiles in real financial markets:
\begin{equation}
\begin{aligned}\label{eqt:Heston model}
\begin{cases}
dS_t = r S_t dt + \sqrt{\nu_t} S_t dW_t^S, \\
d\nu_t  = \kappa (\vartheta - \nu_t) dt + \xi \sqrt{\nu_t}dW_t^\nu,
\end{cases} dW_t^S dW_t^\nu=\rho dt.
\end{aligned}
\end{equation}
Here, $\vartheta$ is the long-term mean of variance, $\kappa$ is the reversion rate, $\xi$ is the volatility of the volatility, and $\rho$ is the correlation of two Brownian motions under the risk-neutral measure. We denote the initial variance as $\nu_{0}$. Hence, it contains five parameters $\theta^{Heston} = \{\nu_0, \vartheta, \kappa, \xi, \rho\}$. 

To comprehensively assess our model, we generate the assumed RND with various levels of skewness by changing two parameters $\xi$ and $\rho$: $\xi=0.35, \rho=-0.9$ for left-skewed RND, $\xi=0.25, \rho=-0.2$ for likely normal RND, and $\xi=0.2, \rho=0.85$ for right-skewed RND, with $\nu_0=0.05, \vartheta=0.25, \kappa=0.15$. We also consider options with a longer time-to-maturity $\tau=2.0$ (2 years) and use left-skewed RND parameters. In the case of a longer time-to-maturity, the option strike range is extended to $[100, 1600]$ to ensure an accurate fit of the RND tails.
These scenarios represent different market conditions, enabling a comprehensive evaluation of our proposed models. Then we train the three models and recover the estimated RNDs to compare with the true Heston RNDs.

\begin{figure}[t]
  \centering
  \includegraphics[width=.95\linewidth]{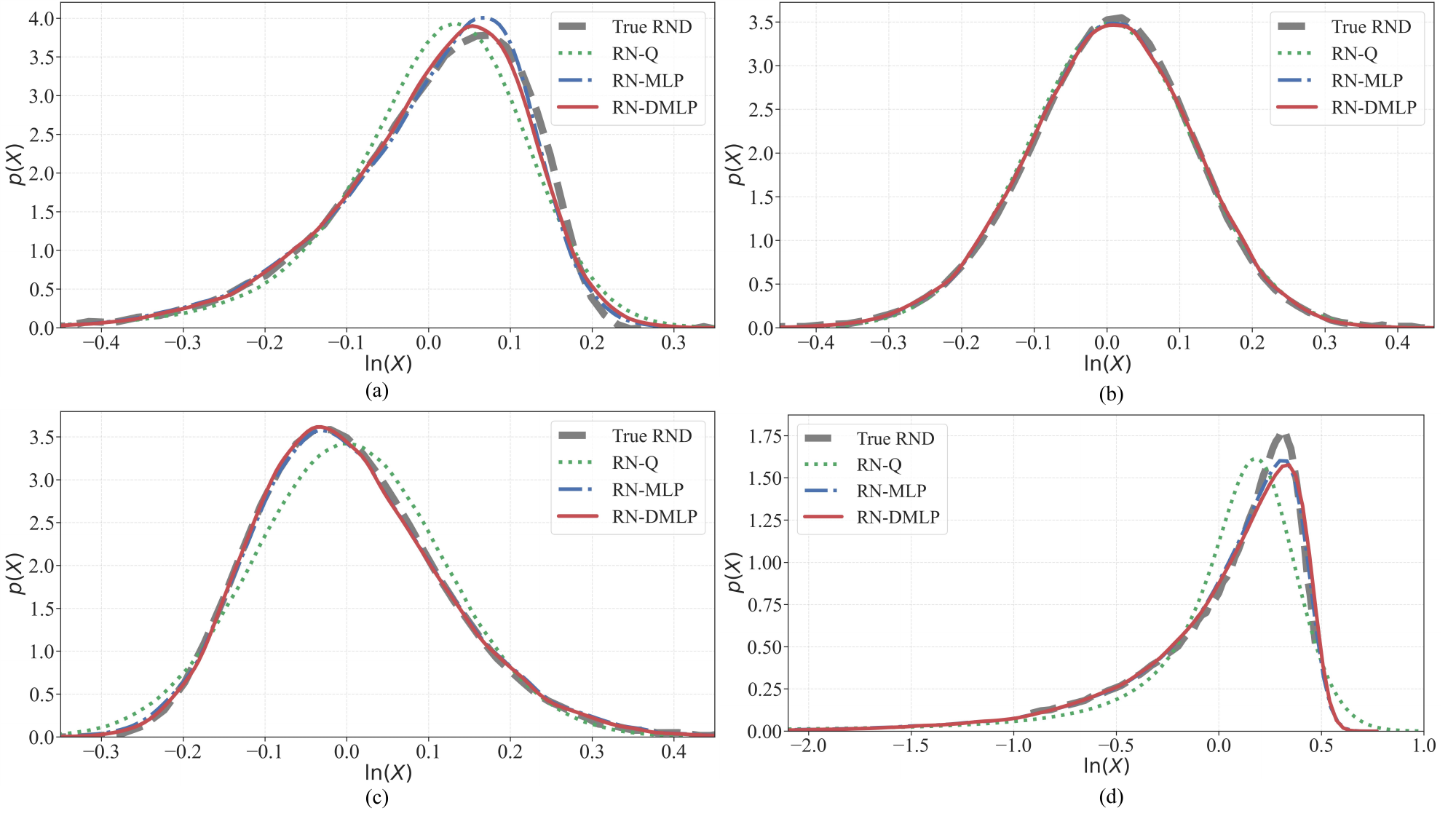}
  \caption{\footnotesize The true Heston RND and the recovered RNDs by our proposed models. Overall, RN-DMLP gives the closest recovered RND. Subfigure (a) displays the left-skewed RND generated by the Heston model with parameters: $\nu_0=0.05, \vartheta=0.25, \kappa=0.15, \xi=0.35, \rho=-0.9$; Subfigure (b) displays the likely-normal RND generated by the Heston model  with parameters: $\nu_0=0.05, \vartheta=0.25, \kappa=0.15, \xi=0.25, \rho=-0.2$; Subfigure (c) displays the right-skewed RND generated by the Heston model with parameters: $\nu_0=0.05, \vartheta=0.25, \kappa=0.15, \xi=0.2, \rho=0.85$. The time-to-maturity $\tau$ in Subfigures (a), (b), and (c) is $0.25$. Subfigure (d) displays the left-skewed RND generated by the Heston model with the parameters in (a) and a longer time-to-maturity $\tau=2.0$.}
  \label{fig:HestonRND}
\end{figure}

\begin{figure}[t]
	\centering
	\includegraphics[width=.95\linewidth]{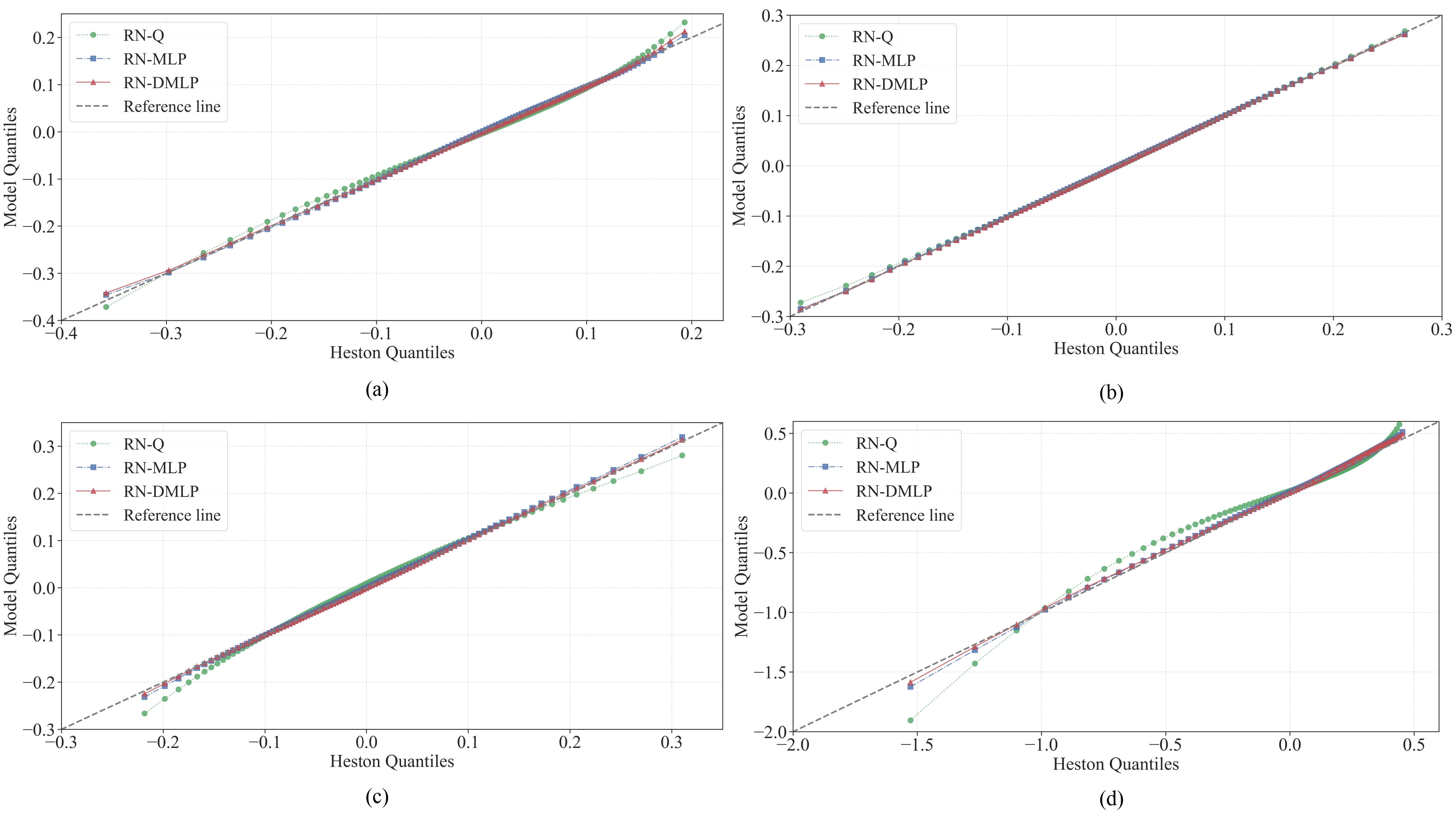}
	\caption{\footnotesize Q-Q plots of risk-neutral densities produced by our models against the Heston RND.}
	\label{fig:HestonQQplot}
\end{figure}

Figure \ref{fig:HestonRND} illustrates the true Heston RND alongside the recovered RNDs by the three models: RN-Q, RN-MLP, and RN-DMLP. Our models demonstrate a notable capability to capture the primary characteristics of the true RND, including main shapes and tail behaviors. Particularly, the RN-DMLP model stands out by exhibiting superior performance. It effectively approximates the true RND and even achieves perfect recovery of the left-skewed or right-skewed RND, surpassing the performance of the other two which show a slight degradation. These simulation outcomes provide robust evidence of the reliability of our proposed models in extracting the RND from option data. 

We further display the Q-Q plots for the four simulation scenarios in Figure \ref{fig:HestonQQplot}, comparing the quantiles of our proposed models against those of the Heston model. The results consistently demonstrate an excellent fit between our models and the theoretical benchmark. The Q-Q plots clearly show that the quantiles of our models align almost perfectly with the quantiles of the Heston model, including tail-side quantiles. This provides direct evidence that our models could replicate the tail behavior of the Heston RND.

\subsection{Empirical Studies with S\&P 500 Options}

In this section, our aim is to compare the empirical performance of our three models with that of classical option pricing models using a real-world dataset. We utilize European option data of the S\&P 500, obtained from the \textit{OptionMetrics IvyDB US} database, accessed through the \textit{Wharton Research Data Services}.


\textit{Data and Preprocessing.}
We collect the daily European option prices and continuous dividend rates of S\&P 500 from \textit{OptionMetrics} for the period spanning from January 4, 1996, to February 28, 2023. Initially, we filter out observations with bid or ask prices lower than \$0.025 to mitigate the impact of decimalization, in line with the approach outlined in \cite{song2016tale}. Additionally, we approximate the option price using the average of the bid and ask (the mid quote). Options with invalid implied volatility are excluded from the dataset. The daily closing price of the index serves as the underlying price in our studies. We normalize the time-to-maturity by dividing the days to maturity by 365. The risk-free interest rate $r$ is linearly interpolated to align with the option maturity. 

\textit{Training, Testing, and Extreme-Moneyness Sets.} 
To compare our models against classical models, at each observation date, we split all options into three disjoint sets.
Except for the training set, we evaluate the pricing performance on both a testing set and an extreme moneyness option set. To achieve this, we arrange strike prices $K$ in ascending order for a given option maturity and select options with moneyness $K/S_t$ falling within the range of $[0.8, 1.2]$ as the training-testing set. We allocate odd-indexed options to the training set and even-indexed options to the testing set. Options outside this moneyness range are categorized as the extreme moneyness option set. The options in the testing set and extreme moneyness set are solely for prediction or out-of-sample evaluation purpose, while the training set is for calibration. 
The training is done via the MSE loss, i.e.,
$L(C_i,\hat{C}_i) = (C_i-\hat{C}_i)^2$ and $L(P_j,\hat{P}_j) = (P_j-\hat{P}_j)^2$.
The out-of-sample prediction/evaluation (testing) is conducted using both MSE and relative MSE, the latter of which is computed by
$L(C_i,\hat{C}_i) = (\hat{C}_i/C_i-1)^2$ and $L(P_j,\hat{P}_j) = (\hat{P}_j/P_j-1)^2$.
The training process involves sampling $Z = [Z_1, Z_2, \dots, Z_N]$ from $\mathbb{N}(0,1)$ with $N=1 \times 10^6$.

\subsection{Two Experimental Settings and Results}
\label{chap_SingleTau}

It is important to note that options on the S\&P 500 display a diverse range of time-to-maturities ($\tau$) on each trading day. Some traditional models, such as the double log-normal \citep{bahra1997implied}, generalized beta  \citep{bookstaber1987general}, and Edgeworth expansion \citep{rubinstein1998edgeworth} models, are generally calibrated for a single $\tau$ value. In contrast, our proposed models represent a significant advancement by allowing multiple $\tau$ values, hence allowing for the modeling of $\tau$-dependent RNDs, forming a term structure of RNDs. To be consistent with those previous models, we investigate two experimental settings: a single $\tau$ setting and a multiple $\tau$ setting.

\begin{figure}[t]
  \centering
  \includegraphics[width=.95\linewidth]{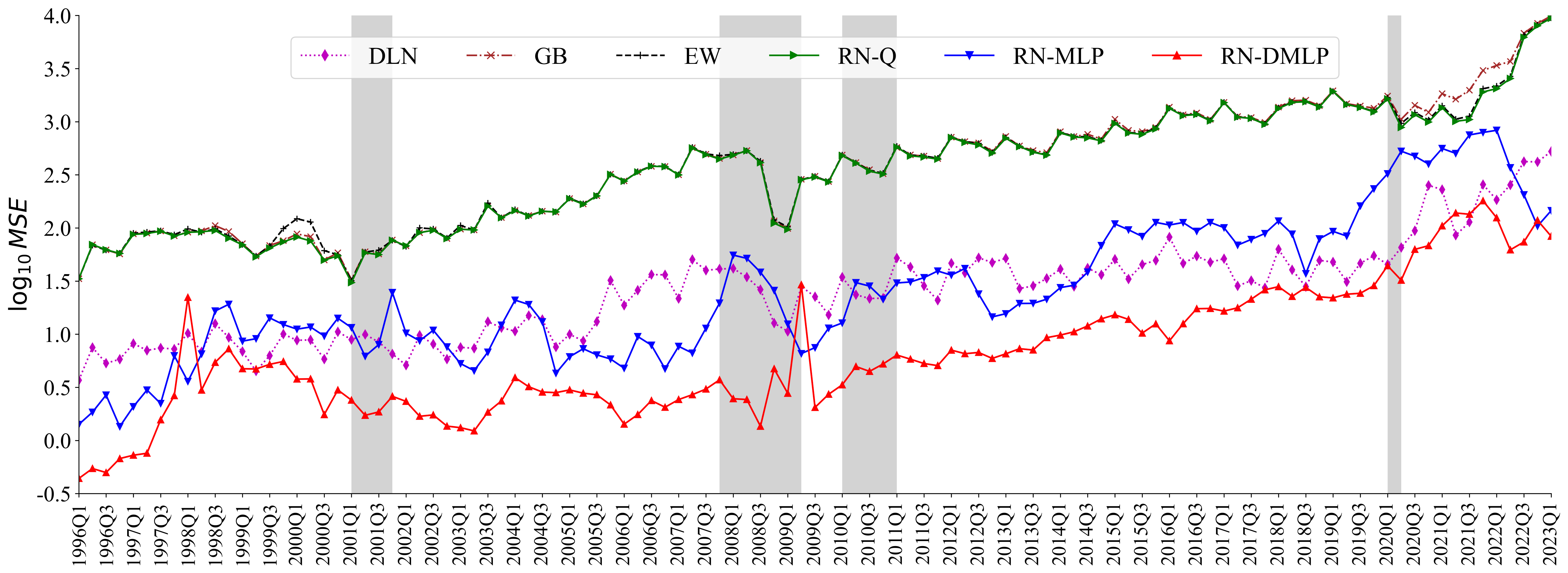}
  \caption{\footnotesize Pricing performance under the single $\tau$ setting. The $\log_{10}\text{MSE}$ (averaged in every quarter) on the testing set is plotted against varying time for each model. The shaded areas represent the financial crisis periods. Overall, RN-DMLP achieves the best performance.}
  \label{fig:dailyMSE_OneTau}
\end{figure}

\begin{table}[t]
  \begin{center}
    \caption{Pricing performance averaged over all years under the single $\tau$ setting.}
    \label{tab:MSEfittedprices_OneTau}
	\begin{tabular}{llccccccccc}
		\toprule
		{} & {} & RN-Q &  RN-MLP &  RN-DMLP & DLN & GB & EW \\
		\midrule
		\multirow{2}{*}{Testing} & {MSE} & 23.96 & \textit{8.36} & \textbf{3.29} & 15.42 & 184.40 & 42.84 \\
		{} & {Relative} & 0.14 & \textit{0.12} & \textbf{0.08} & 0.15 & 0.29 & 0.51 \\
		\multirow{2}{*}{Extreme} & {MSE} & 110.78 & \textit{53.86} & \textbf{43.55} & 164.95 & 430.49 & 231.02 \\
		{} & {Relative} & 1.35 & \textit{0.46} & \textbf{0.39} & 1.88 & 1.64 & 4.45 \\
		\bottomrule
	\end{tabular}

  \medskip
  {\raggedright \textit{Notes.} The smallest loss is highlighted in bold, and the second smallest is in italics. We present the results on both the testing set and the extreme moneyness set, under the metrics of both MSE and relative MSE. Consistently, RN-DMLP achieves the best performance. \par}
  \end{center}
\end{table}

\paragraph{Single Time-to-maturity Setting}
In our first experimental setting, the models are calibrated based on option contracts with each specific time-to-maturity ($\tau$) separately. Given the presence of different maturities on each trading day, this approach yields multiple calibrated models on a daily basis. This experimental setup is consistent with standard practices commonly employed in the literature. We benchmark our models, RN-Q, RN-MLP, and RN-DMLP, against three classical models: the double log-normal (DLN), the generalized beta (GB), and the Edgeworth expansion (EW). Our models are calibrated following the default configurations described in Section \ref{par:hyperparameter}, while the benchmark models are calibrated using the R Package \textit{Risk-neutral Density Extraction} \citep{hamidieh2014rnd}. Subsequently, these calibrated models are utilized to predict option prices in both the testing set and the extreme-moneyness set. We present the mean squared error (MSE) and relative MSE to evaluate the out-of-sample pricing performance and extreme-moneyness pricing capabilities of these models.

The empirical evidence presented in Figure \ref{fig:dailyMSE_OneTau} clearly demonstrates the superior performance of our RN-DMLP model, as it consistently exhibits lower MSE levels compared to alternative models across years from 1996 to 2023. Specifically, the MSE of the RN-DMLP model remains below $10^1$ most of the time and does not exceed $10^3$, indicating a significant improvement over its competitors. Our RN-MLP model displays comparable accuracy to the DLN method. Table \ref{tab:MSEfittedprices_OneTau} further corroborates that our proposed RN-DMLP model consistently achieves lower MSE and relative MSE on both the testing set and the extreme-moneyness set compared to other baseline models. The DLN method shows acceptable performance on the testing set, but exhibits noticeably lower pricing accuracy when dealing with extreme-moneyness options. Overall, our RN-DMLP model demonstrates superior predictive accuracy on both sets compared to all other models.

\paragraph{Multiple Time-to-maturities Setting}
In the subsequent analysis, we delve into the multiple $\tau$ setting. This involves calibrating a single model using multiple option contracts with a range of time-to-maturities $\tau$ on a trading day. We continue to use the same setting to divide and obtain the training set, testing set, and extreme-moneyness set. To better guarantee the no-arbitrage Constraint \ref{C5} and \ref{C6}, we establish a dense grid of time-to-maturity $\tau$ and strike price $K$, to be used in the penalty term in Eqn. \eqref{eq:objective function-DMLP}. Suppose the options from the market on a trading day have $N_\tau$ maturities $\{\tau_i\}_{i=1}^{N_\tau}$ and $N_K$ strike prices $\{K_i\}_{i=1}^{N_K}$. The synthetic grid $\mathcal{I}^{syn}$ is given as follows:
\begin{equation}
\begin{aligned}\label{eq:grid}
\mathcal{I}^{syn} = \biggl\{ (\tau^{syn},K^{syn}): & \tau^{syn} \in \left\{\frac{\tau_{[i]} + \tau_{[i+1]}}{2} \right\} \cup \{\tau_i\},\\
 & K^{syn} \in \left\{\frac{K_{[i]} + K_{[i+1]}}{2}\right\} \cup \{K_i\} \biggl\}, 
\end{aligned}
\end{equation}
where $\tau_{[i]}$ or $K_{[i]}$ is the $i$-th element in ascending order in the set $\{\tau_i\}$ or $\{K_i\}$. We set about $1,000$ points in $\mathcal{I}^{syn}$ in total on each training day for training our models. 

Subsequently, we calibrate our models with option data, incorporating multiple $\tau$ values as inputs. To provide a comprehensive and reliable evaluation of the models' performance in this setting, we conduct extensive comparisons involving a total of 9 classical models and their variants: variance gamma (VG) \citep{madan1990variance}, variance gamma with Cox–Ingersoll–Ross process (VGCIR) \citep{schoutens2003pricing}, variance gamma with Gamma Ornstein–Uhlenbeck clock process (VGGOU) \citep{carr2003stochastic}, normal inverse Gaussian (NIG) \citep{barndorff1997normal}, normal inverse Gaussian with Cox–Ingersoll–Ross process (NIGCIR) \citep{carr2003stochastic}, normal inverse Gaussian with Gamma Ornstein–Uhlenbeck clock process (NIGGOU) \citep{eberlein1999term}, Merton with jumps (MJ) \citep{merton1976option}, Heston method \citep{heston1993closed}, and Bates method \citep{bates1996jumps}. These models will be calibrated using MATLAB packages provided by \cite{kienitz2013financial}.

\begin{figure}[t]
  \centering
  \includegraphics[width=.85\linewidth]{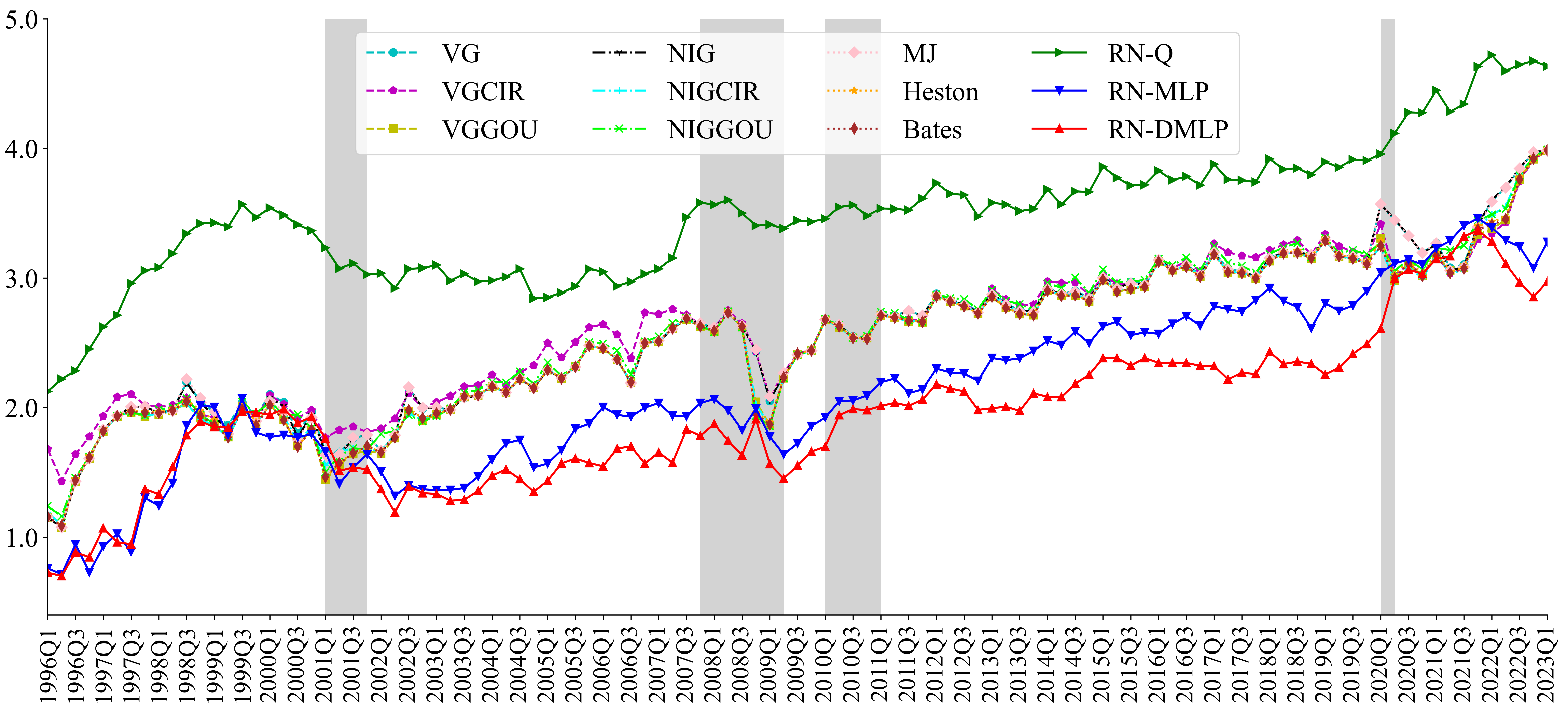}
  \caption{\footnotesize Pricing performance under the multiple $\tau$ setting. The $\log_{10}\text{MSE}$ (averaged in every quarter) on the testing set is plotted against varying time for each model. The shaded areas represent the financial crisis periods. Overall, RN-DMLP achieves the best performance and RN-MLP is the second best.}
  \label{fig:dailyMSE_MulTaus}
\end{figure}

\begin{table}[t]
  \begin{center}
    \caption{Pricing performance averaged over all years under the multiple $\tau$ setting.}
    \label{tab:MSEfittedprices_MulTaus}
    \resizebox{\textwidth}{!}{
	\begin{tabular}{llcccccccccccc}
	\toprule
	{} & {}  & RN-Q & RN-MLP & RN-DMLP & VG & VGCIR & VGGOU & NIG & NIGCIR & NIGGOU & MJ & Heston & Bates \\
	\midrule
	\multirow{2}{*}{Testing} & {MSE} & 9263.23 & \textit{546.69} & \textbf{348.64} & 1286.16 & 1149.08 & 1079.02 & 1285.80 & 1127.50 & 1204.06 & 1279.51 & 1107.22 & 1085.15 \\
	{}  & {Relative} & 4472.28 & 3.49 & \textbf{0.43} & 2.78 & 5.87 & \textit{1.06} & 2.21 & 4.05 & 13.21 & 1.73 & 5.11 & 2.86 \\
	\multirow{2}{*}{Extreme} & {MSE} & 1742.48 & \textit{562.49} & \textbf{391.10} & 1480.01 & 1401.38 & 1440.08 & 1502.32 & 1718.70 & 1460.80 & 1543.71 & 2371.34 & 1959.88 \\
	{} & {Relative} & 2277.36 & \textit{4.98} & \textbf{1.11} & 4.47 & 15.89 & 8.51 & 5.16 & 29.61 & 19.30 & 12.18 & 1989.34 & 1711.25 \\
	\bottomrule
	\end{tabular}
	}

\medskip
{\raggedright \textit{Notes.}  The smallest loss is highlighted in bold, and the second smallest is in italics. We present the results on both the testing set and the extreme-moneyness set, under the metrics of both MSE and relative MSE. Overall, RN-DMLP achieves the best performance and RN-MLP is the second best. \par}
  \end{center}
\end{table}

Figure \ref{fig:dailyMSE_MulTaus} reveals that our models RN-MLP and RN-DMLP incorporating neural networks achieve a substantial reduction in MSE compared to the parametric method RN-Q. They also demonstrate evident superiority to the 9 classical methods, achieving MSE levels in the order of approximately $10^2$ most of the time. Table \ref{tab:MSEfittedprices_MulTaus} further substantiates the superior performance of our RN-DMLP model, as it consistently obtains lower MSE and relative MSE on both the testing set and the extreme-moneyness set. Notably, RN-DMLP attains nearly an order of magnitude reduction in MSE compared to the classical competitors. Our RN-MLP model also exhibits lower MSE and relative MSE, especially when pricing the extreme-moneyness options when compared to the classical models. This table further indicates that the classical models exhibit similar performance levels. Both Figure \ref{fig:dailyMSE_MulTaus} and Table \ref{tab:MSEfittedprices_MulTaus} clearly showcase the significant superiority of our proposed models.

To further strengthen the empirical studies, we complement additional benchmarks with neural-network-based models and place the comparison results in Appendix \ref{appen:deep}.
Moreover, in Appendix \ref{appen:violation}, we conduct two sets of additional experiments to (1) quantify how strongly the condition $J^{\mu} = 0$ is violated in practice, and (2) investigate whether alternative drift specifications—specifically a flexible, learnable drift or a hard constraint enforcing—can improve calibration.

\subsection{Testing the Stability of Extracted RND}
\label{chap:StabilityExp}

In this section, we aim to conduct an experiment to assess the stability of the RND extracted by our proposed models and a total of 12 competitors previously mentioned. For this purpose, we randomly select S\&P 500 option data on a specific trading day (August 15, 2019) with a specific time-to-maturity ($\tau=$ 400 days) to perform the experiment. To assess the RND stability, we introduce small perturbations to the observed option data. These small perturbations are equivalent to one tick size of the S\&P 500 option prices, which is \$0.25. Every option price in the training set will randomly obtain either an addition or a subtraction of one tick. We then re-estimate the RND using our models and the twelve competitors. This procedure is repeated 50 times, allowing us to measure the deviations in perturbed RNDs.

Given the subtle nature of these perturbations, the resultant changes in the RNDs should be negligible. Models that exhibit smaller shifts in distribution in response to these perturbations are considered to have superior stability. However, the direct comparison of a large number of probability density functions (PDFs) poses a significant analytical challenge. To address this, we focus our evaluation on the variations in the distributional characteristics, specifically on the standard deviations of ten PDF characteristics across the fifty trials of perturbations. These standard deviations serve as our indicators of RND stability. The ten PDF characteristics are: the mean $\mu$, the standard deviation $\sigma$, the skewness, Pearson median-based skewness defined as $\text{Skew}_{\text{PM}} = \frac{\mu-X_{50}}{\sigma}$ where $X_{50}$ is the 50\%-quantile, the asymmetry measure defined as $\text{Skew}_{\text{AM}}=\frac{X_{75}-X_{50}}{X_{50}-X_{25}}$, the kurtosis, and four tail-side percentiles $X_{01}, X_{05}, X_{95}, X_{99}$.

\begin{table*}[t]
  \caption{The standard deviations of the RND characteristics across fifty trials of perturbations as estimated by various models.}
  \label{tab:std_PerturbRND}
  \resizebox{\linewidth}{!}{
	\begin{tabular}{llllllllllllllll}
	\toprule
	& RN-Q & RN-MLP & RN-DMLP & DLN & GB & EW & VG & VGCIR & VGGOU & NIG & NIGCIR & NIGGOU & MJ & Heston & Bates \\
	\midrule
	MSE & 122.458 & 12.607 & 2.339 & 46.598 & 111.919 & 33.215 & 69.593 & 68.316 & 166.051 & 32.342 & 48.948 & 32.868 & 139.630 & 34.284 & 33.899 \\
	$\mu$ & $3.06\times10^{-5}$ & 0.002 & $3.51\times10^{-5}$ & 229.834 & 0.838 & 1.264 & $1.73\times10^{-5}$ & 0.025 & 9.055 & $6.68\times10^{-4}$ & 0.346 & 0.039 & $1.39\times10^{-4}$ & 0.041 & 0.033 \\
	$\sigma$ & $3.49\times10^{-4}$ & 0.003 & $9.68\times10^{-5}$ & 52.372 & 0.417 & 1.159 & $4.56\times10^{-5}$ & 0.236 & 50.380 & 0.003 & 2.896 & 0.306 & $5.92\times10^{-4}$ & 0.187 & 0.102 \\
	Skewness & 0.042 & 0.019 & 0.002 & 0.200 & $6.44\times10^{-4}$ & 0.001 & 0.003 & 0.038 & 3.299 & 0.571 & 0.263 & 0.028 & 0.012 & 0.044 & 0.045 \\
	$\text{Skew}_{\text{PM}}$ & $1.43\times10^{-5}$ & 0.004 & $4.31\times10^{-4}$ & 0.137 & 0.152 & 0.158 & $6.22\times10^{-4}$ & 0.003 & 0.143 & 0.310 & 0.026 & 0.005 & 0.005 & 0.002 & 0.002 \\
	$\text{Skew}_{\text{AM}}$ & $1.17\times10^{-4}$ & 0.008 & $9.83\times10^{-4}$ & 0.167 & 0.167 & 0.167 & $5.03\times10^{-11}$ & 0.018 & 0.259 & $5.45\times10^{-11}$ & 0.073 & 0.017 & $4.76\times10^{-11}$ & 0.017 & 0.017 \\
	Kurtosis & 4.732 & 0.107 & 0.015 & 0.834 & 0.002 & 0.006 & 0.047 & 0.218 & 36.366 & 12.042 & 3.202 & 0.230 & 0.040 & 0.222 & 0.231 \\
	$X_{01}$ & $9.21\times10^{-4}$ & 0.008 & $7.83\times10^{-4}$ & 3.846 & 3.846 & 3.846 & $9.68\times10^{-5}$ & 2.522 & 62.457 & 0.026 & 23.991 & 2.787 & $4.28\times10^{-4}$ & 2.506 & 2.473 \\
	$X_{05}$ & $1.57\times10^{-4}$ & 0.007 & $2.28\times10^{-4}$ & 22.992 & 22.992 & 22.992 & $1.70\times10^{-5}$ & 0.439 & 29.062 & $6.23\times10^{-4}$ & 0.658 & 0.450 & 0.001 & 0.467 & 0.221 \\
	$X_{95}$ & $5.54\times10^{-5}$ & 0.001 & $1.71\times10^{-4}$ & 37.241 & 37.241 & 37.241 & $1.00\times10^{-6}$ & 0.686 & 26.936 & $3.28\times10^{-5}$ & 2.179 & 0.216 & $1.12\times10^{-4}$ & 1.011 & 0.970 \\
	$X_{99}$ & $8.40\times10^{-5}$ & 0.001 & $3.29\times10^{-4}$ & 18.286 & 18.286 & 18.286 & $1.24\times10^{-6}$ & 2.432 & 122.421 & $6.55\times10^{-6}$ & 0.582 & 0.335 & 0.005 & 3.198 & 3.011 \\
	\bottomrule
	\end{tabular}}

\medskip
{\raggedright \textit{Notes.}   Lower values in the table indicate superior stability of the estimated RND. The first row (MSE) represents the average mean square pricing error across the fifty trials. It is unequivocal that the RN-DMLP model exhibits the highest level of stability. \par}
\end{table*}

\begin{figure}[t]
  	\centering
  	\includegraphics[width=\linewidth]{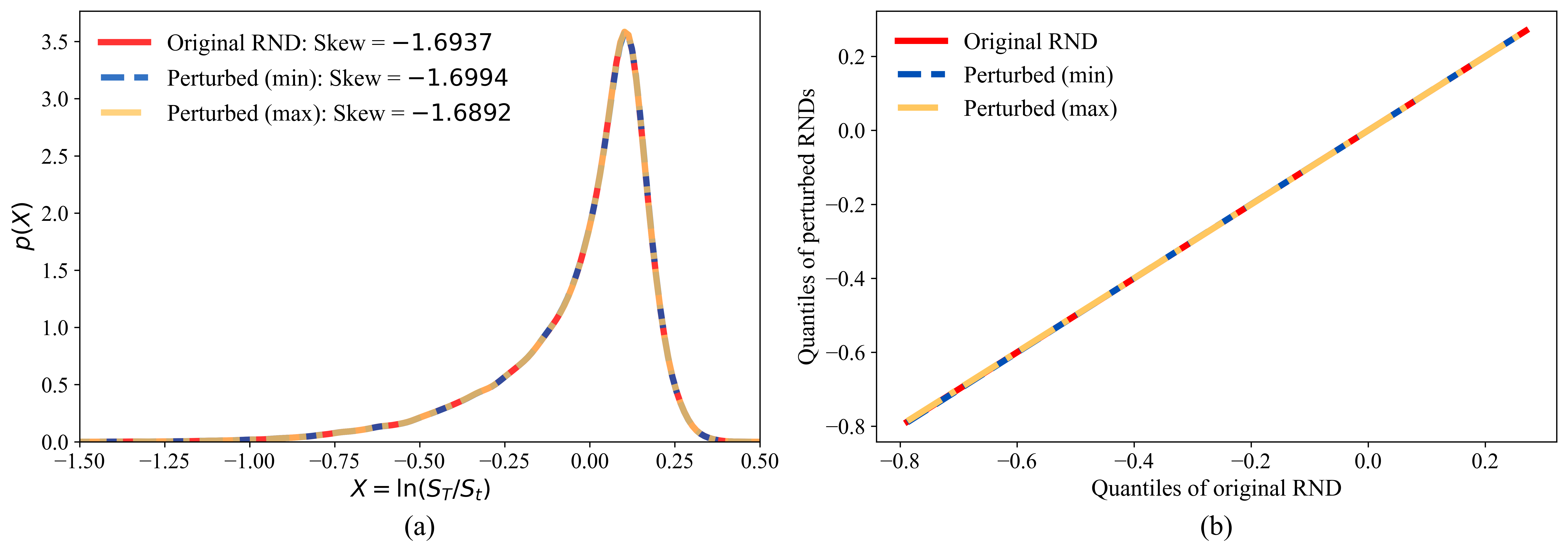}
	\caption{\footnotesize 
			(a) Probability density functions and (b) Q-Q plots for original and perturbed risk-neutral densities. 
			The original RND (red, skewness = $-1.6937$) is compared with its minimum-skewness perturbation (blue, $-1.6994$) and maximum-skewness perturbation (yellow, $-1.6892$), showing that our model RN-DMLP remains robust under moderate price perturbations.
	}
  \label{fig:PerturbRND}
\end{figure}

Table \ref{tab:std_PerturbRND} reports the standard deviations of the RND characteristics across fifty perturbation trials. The results provide strong quantitative evidence of the superior robustness of our neural-based models, particularly the RN-DMLP. The RN-DMLP demonstrates the most remarkable overall stability among those competitors. The RN-MLP performs comparably, showing similar stability yet with slightly larger fluctuations than RN-DMLP, while the RN-Q exhibits a greater standard deviation in kurtosis despite maintaining a low overall dispersion. Among these competitors, the distributional models (DLN, GB, and EW) demonstrate relatively stable higher-order moments but suffer from large dispersions in the mean and variance. The nine methods that incorporate stochastic processes remain relatively stable in most RND characteristics, albeit occasionally exhibiting larger dispersions or higher pricing errors.

In the first row of Table \ref{tab:std_PerturbRND}, we present the average MSE across the fifty trials of perturbations. Our proposed model, RN-DMLP, not only achieves the most accurate predictions but also ensures the highest stability, as evidenced by the lowest MSE and minimal changes in the extracted RNDs. Figure \ref{fig:PerturbRND} presents an example of perturbations: (a) shows the original RND extracted by RN-DMLP along with the RNDs exhibiting the minimum and maximum skewness across the fifty trials; (b) depicts the Q-Q plots comparing the original RND with these two perturbed RNDs. We observe that the distributions are nearly indistinguishable, with the two perturbed RNDs being highly consistent with the original one, further validating the stability of our method.

\section{Higher-Order Moments of Extracted RNDs}
\label{sec:HigherOrderMoments}

\begin{table}[t]
  \centering
  \caption{The ratio of left and right skewed RNDs extracted by the model RN-DMLP.}
  \label{tab:RatioLeftRightRNDs}
  \begin{tabular}{lccc}
    \toprule
    {} &  Left-skewed &  Right-skewed & Total\\
    \midrule
    Number of RNDs &  64383 (98.72\%) &  833 (1.28\%) & 65216 \\
    \bottomrule
  \end{tabular}
\end{table}

\begin{figure}[t]
  \centering
  \includegraphics[width=\linewidth]{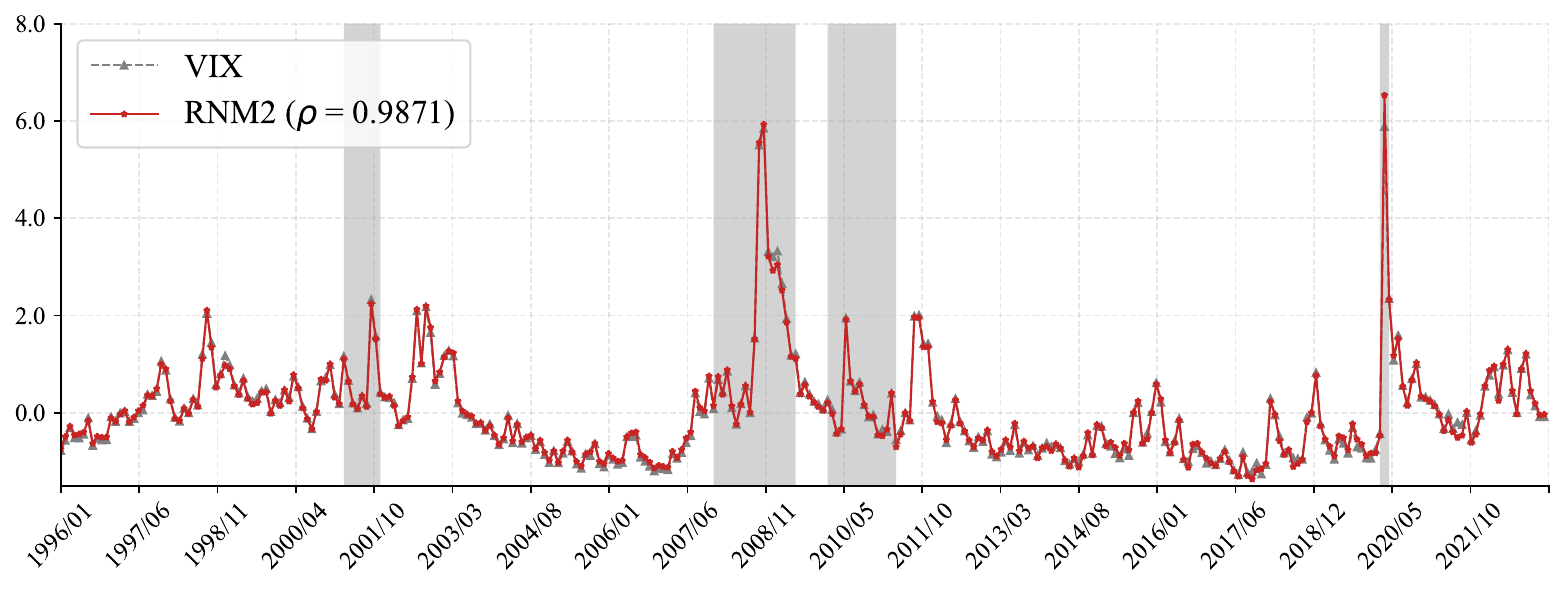}
  \caption{\footnotesize This figure displays the monthly risk-neutral volatility (RNM2) extracted by the RN-DMLP model and the VIX index from January 1996 to February 2023, as well as their correlation coefficient. The RNM2 is computed from the RND extracted using option contracts with time-to-maturities between 25 and 35 days. The shaded grey areas represent the financial crisis periods. In the figure, both series have been standardized to have zero mean and unit variance.}
  \label{fig:RNDMoments_VIX}
\end{figure}

\begin{figure}[t]
	\centering 
	\subfigure[]{\includegraphics[width=0.7\linewidth]{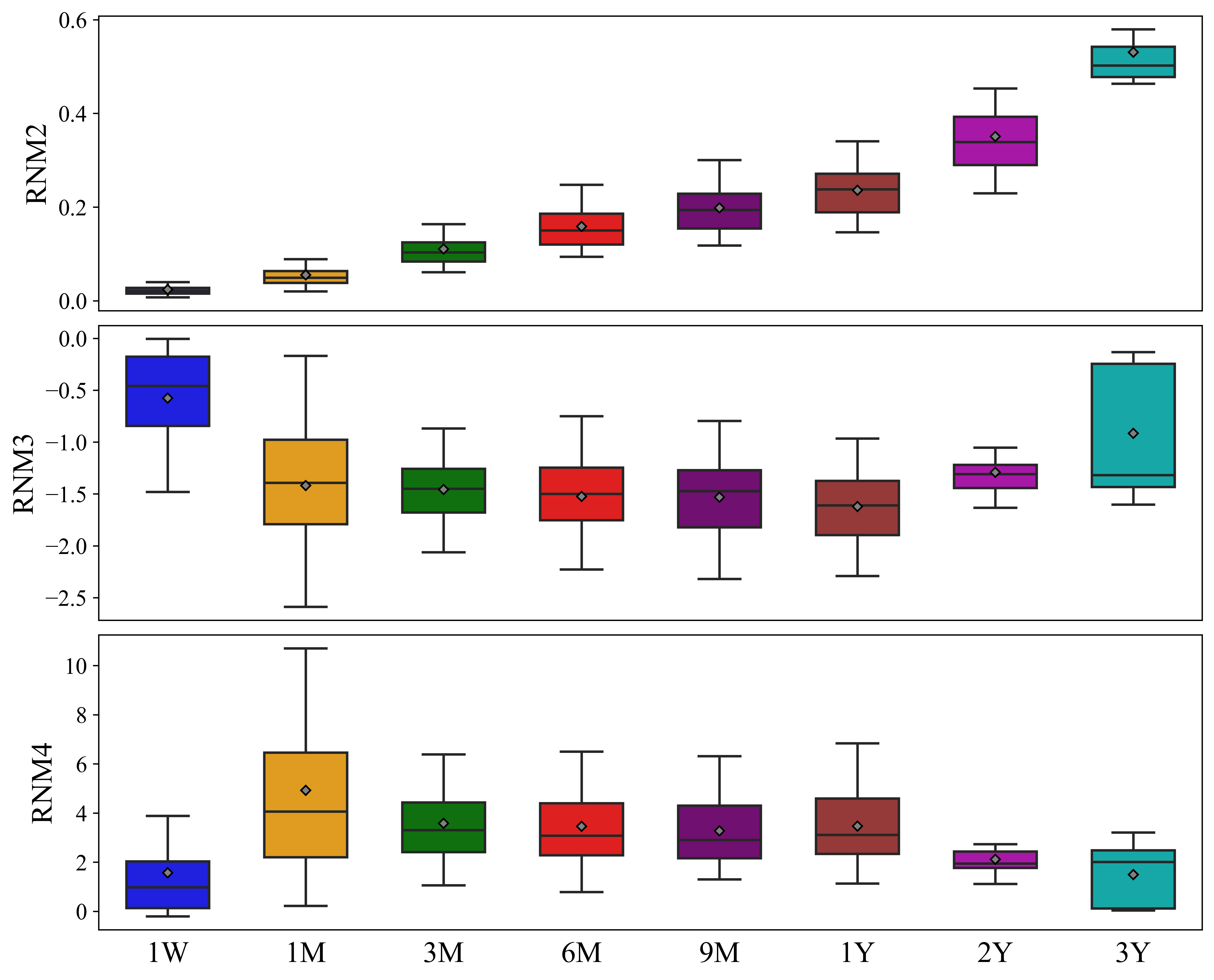}}
	\subfigure[]{\includegraphics[width=\linewidth]{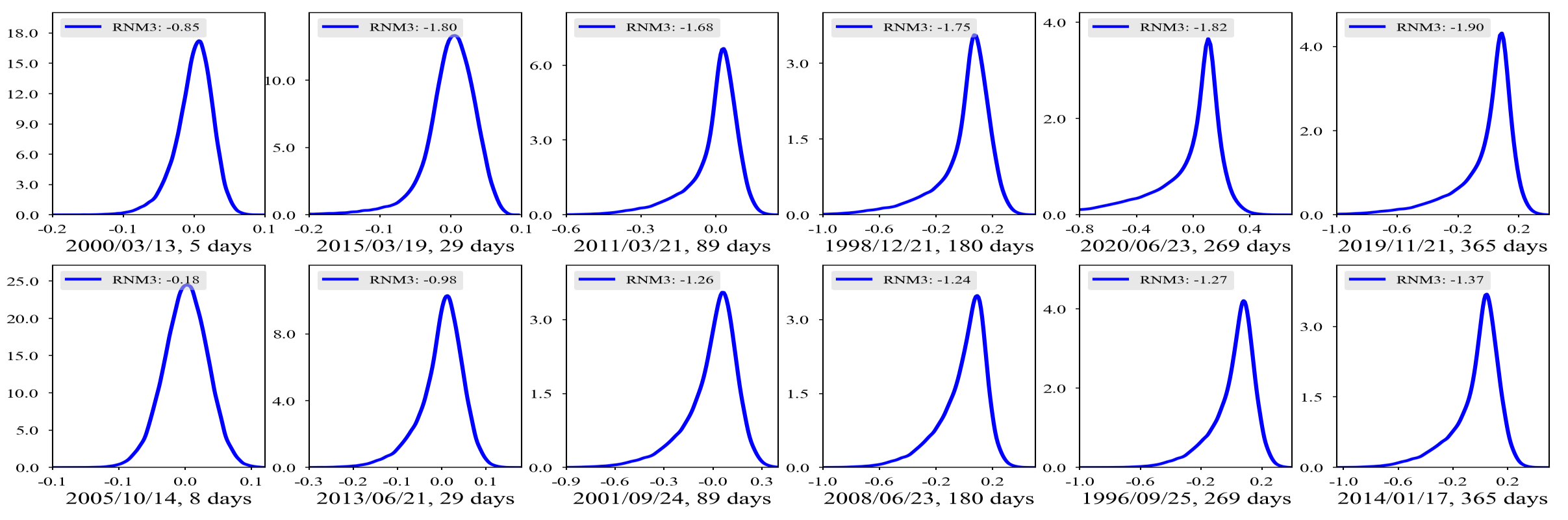}}
	\caption{\footnotesize (a) Box plots of the three risk-neutral moments RNM2 -- RNM4 in every time-to-maturity group ranging from 1 week (1W) to 3 years (3Y).
	(b) Two representative RNDs within each time-to-maturity group, with each column corresponding to a maturity group (1W, 1M, 3M, 6M, 9M, and 1Y).
	In the first row of (b), the RNDs have a skewness level corresponding to the 25th-percentile of all skewness levels within the maturity group. In the second row, the RNDs have a skewness level corresponding to the 75th-percentile.
	}
	\label{fig:RNDStatCurves}
\end{figure}

\begin{figure}[t]
	\centering 
	\subfigure[Risk-Neutral Volatility]{\includegraphics[width=.45\linewidth]{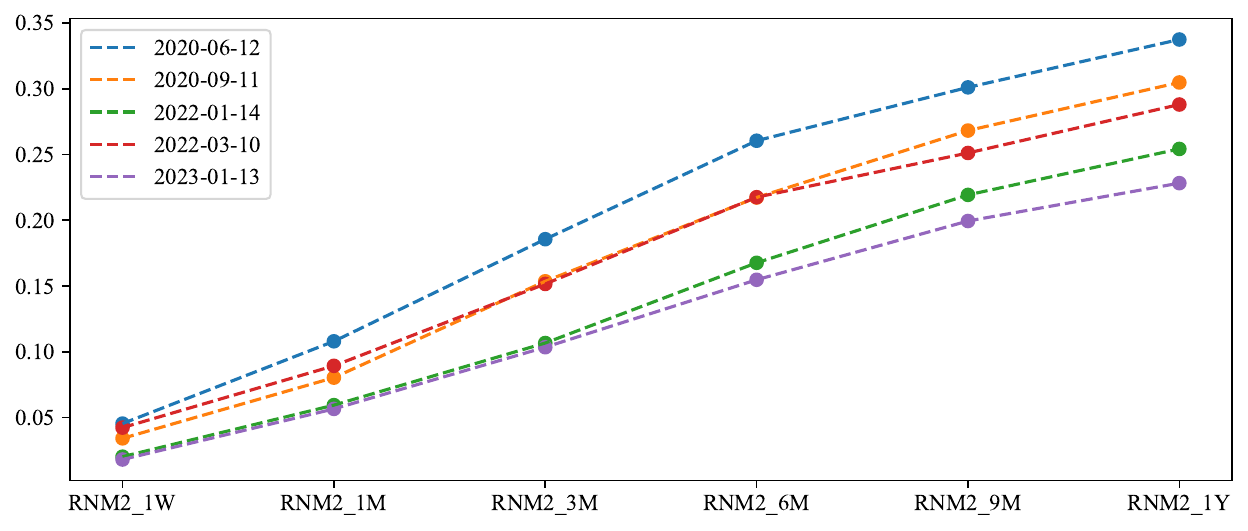}}
	\subfigure[Risk-Neutral Volatility]{\includegraphics[width=.45\linewidth]{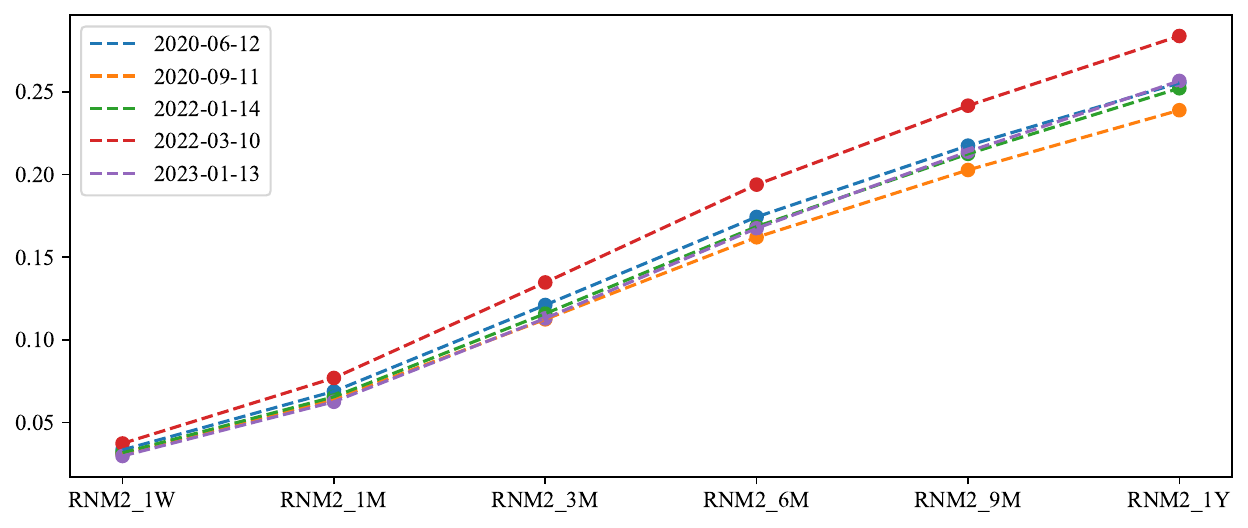}}\\
	\subfigure[Risk-Neutral Skewness]{\includegraphics[width=.45\linewidth]{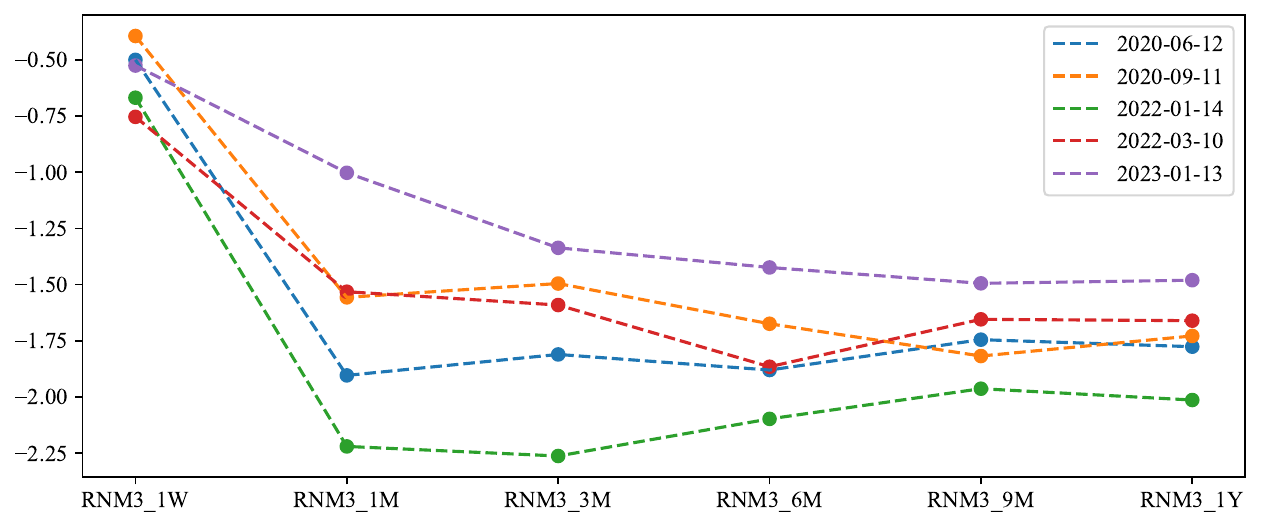}}
	\subfigure[Risk-Neutral Skewness]{\includegraphics[width=.45\linewidth]{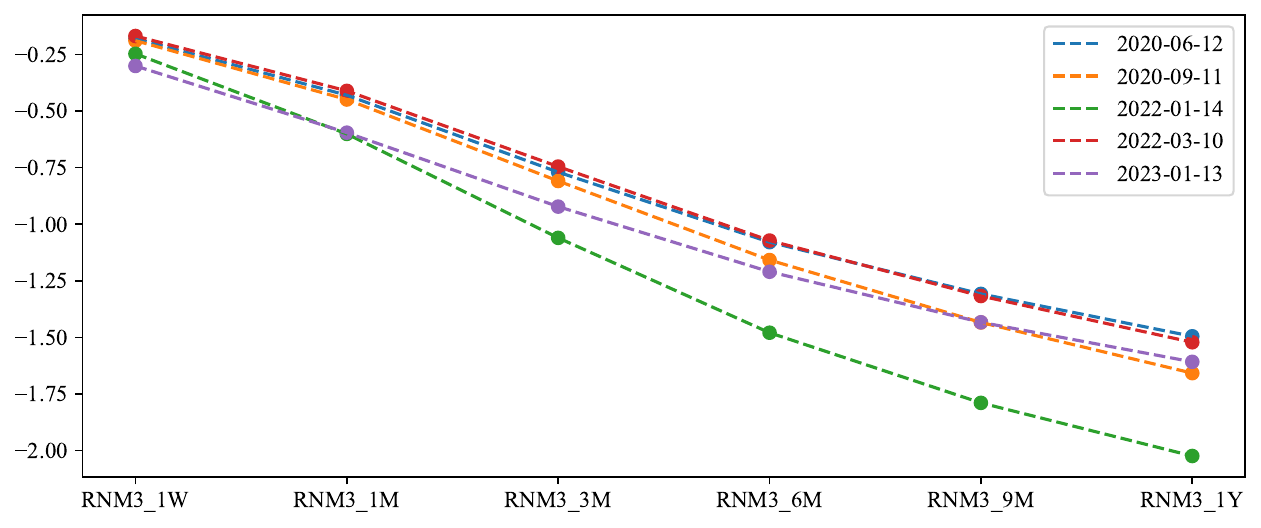}}\\
	\subfigure[Risk-Neutral Kurtosis]{\includegraphics[width=.45\linewidth]{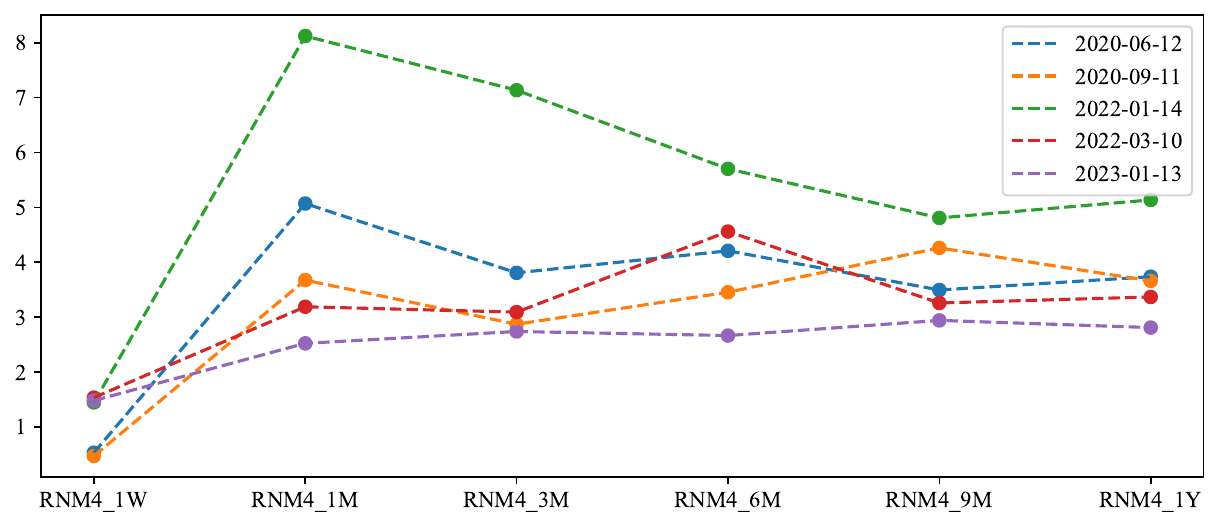}}
	\subfigure[Risk-Neutral Kurtosis]{\includegraphics[width=.45\linewidth]{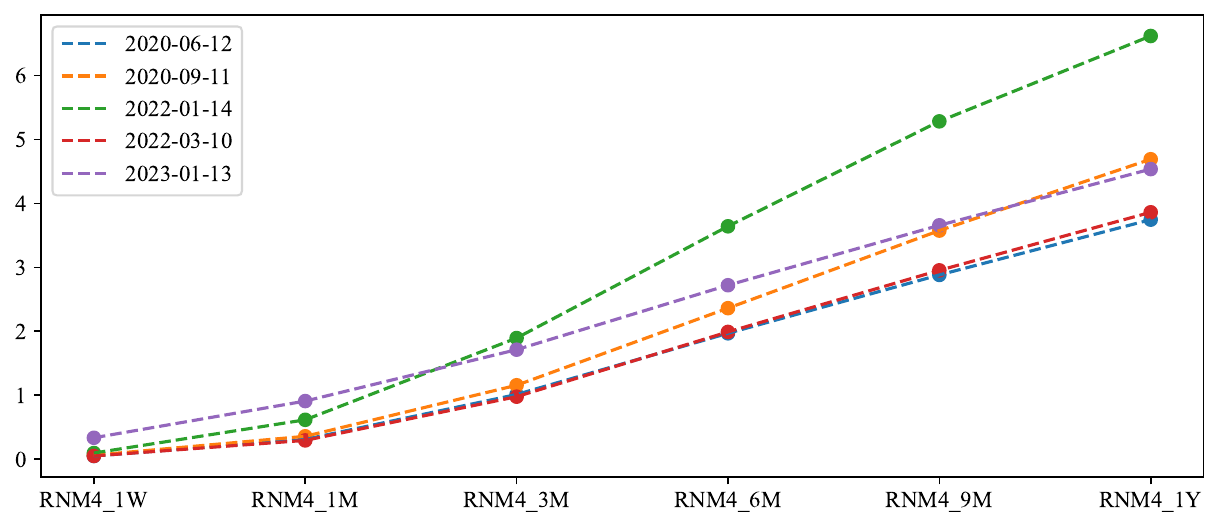}}\\
	\caption{\footnotesize The term structures of risk-neutral moments (RNM2 -- RNM4) produced by our RN-DMLP model (a, c, and e) and by the Heston model (b, d, and f), with the time horizon increasing from 1 week to 1 year. We select five trading days for both RN-DMLP and the Heston. 
	The term structures of RNM3 and RNM4 produced by the RN-DMLP exhibit a variety of patterns across dates, including strictly monotonic profiles, nearly flat profiles, and reverse trends. By contrast, we observe that the Heston model tends to produce smoother, more uniformly monotonic term structures for skewness and kurtosis. For RNM2, both models consistently yield increasing term structures with maturity.
	}
	\label{fig:term_structures}
\end{figure}

In this section, our aim is to derive valuable insights about the RNDs of S\&P 500 \emph{returns} extracted by the RN-DMLP model (not S\&P 500 \emph{prices}). As detailed in Section \ref{chap_SingleTau} and \ref{chap:StabilityExp}, we conducted empirical experiments to rigorously evaluate the option pricing performance. To commence our investigation here, we collect the extracted RNDs from the well-calibrated RN-DMLP models. The RND in Eqn. \eqref{eq:function-X-DMLP} is in a generative machine learning form, allowing us to easily compute the risk-neutral moments as well as density values through sampling. We examine the risk-neutral moments and density graphs of RNDs to show what implications they have. Furthermore, we conduct a comparative analysis with the monthly VIX index, which is a well-established and widely utilized risk measure based on S\&P 500. The sample period we consider spans from January 1996 to February 2023. In the following, we present four key findings. 

\textit{Firstly}, throughout the entire sample period from January 1996 to February 2023, there are 65,216 RNDs in total. Table \ref{tab:RatioLeftRightRNDs} reports that around 98.72\% of cases are left-skewed while around 1.28\% of cases are right-skewed. It indicates the dominance of left-skewed RNDs in the market.

\textit{Secondly}, we display the monthly risk-neutral volatility series and the time series of VIX index in Figure \ref{fig:RNDMoments_VIX}. The risk-neutral volatility is computed from the RND extracted using option contracts with time-to-maturities between 25 and 35 days. As shown in the figure, the two series coincide with each other and the correlation coefficient is 0.9871. These two findings not only align with intuitions on the financial markets, but also verify the rationality and robustness of the extracted RNDs from our model. 


\textit{Thirdly}, for various time-to-maturities (one week, one month, three months, six months, nine months, one year, two years, and three years), we summarize the risk-neutral moments (RNMs) via boxplotting and illustrate some representative extracted RNDs in Figure \ref{fig:RNDStatCurves}. Subfigure~(a) presents the box plots of the three risk-neutral moments RNM2--RNM4 as the time-to-maturity increases from 1W to 3Y (RNM3 and RNM4 are defined as the centered and scaled versions, with an additional
adjustment of subtracting 3 for RNM4).
We observe that the second moment (RNM2) rises progressively as the time-to-maturity increases, while the absolute magnitude of skewness (RNM3) initially rises and then decreases. The kurtosis (RNM4) exhibits a similar pattern as skewness. For longer time-to-maturities (beyond 1 year), both RNM3 and RNM4 decrease towards zero, indicating that the risk-neutral return distribution becomes less heavy-tailed and converges to a Gaussian distribution as time aggregation effects take hold, which is consistent with standard expectations.
Subfigure (b) exhibits two representative RNDs within each time-to-maturity group, with each column corresponding to one group. In the first row of (b), the RNDs have a skewness level corresponding to the 25th-percentile of all skewness levels within the maturity group. In the second row, the RNDs have a skewness level corresponding to the 75th-percentile.


\textit{Finally}, we investigate the term structures of the RNMs generated by the RN-DMLP model on each trading day and compare them with those obtained from Heston model. Figure \ref{fig:term_structures} plots RNM2 -- RNM4 for time-to-maturities from one week to one year. RNM3 and RNM4 produced by the RN-DMLP exhibit a variety of patterns across dates, including strictly monotonic profiles, nearly flat profiles, and reverse trends. By contrast, we observe that the Heston model tends to produce smoother, more uniformly monotonic term structures for skewness and kurtosis. For RNM2, both models consistently yield increasing term structures with maturity.
These findings suggest that the RN-DMLP's flexible parameterization allows it to capture a wider variety of skewness and kurtosis term-structure patterns than the Heston model.

\section{Conclusion}\label{sec:Conclusion}

Extracting the risk-neutral density (RND) is a critical and challenging task in finance. In this paper, we introduce the risk-neutral generative network (RNGN) framework, featuring three generative models that do not rely on assumptions regarding the underlying asset price dynamics or the distributional families to which the RND belongs. Our proposed models adhere to the no-arbitrage constraints. In simulation studies, we generate option price data encompassing a range of RNDs with various shapes (including left-skewed, nearly normal, and right-skewed) to demonstrate that our three models can efficiently extract the true RND. In the empirical analysis, we compare the results obtained under our proposed framework with those derived from classical methodologies, utilizing option data of S\&P 500. Empirical results underscore the superiority of our models in terms of pricing accuracy on both testing sets and extreme-moneyness sets. Our models further exhibit greater stability of the estimated RND. It is noteworthy that the RN-DMLP model performs exceptionally well, achieving an extremely low MSE level.

Our model RN-MLP incorporates a neural network known as Multilayer Perceptron (MLP), based on the structure of our initial model RN-Q. Our derived pricing formulas are based on Monte Carlo simulation, and rational considerations are taken into account when learning our network. We attribute the exceptional performance of RN-MLP to the learning capability of the neural network, which enables it to capture complex patterns in the data. Inspired by the idea of double log-normal density, we further combine two RN-MLPs to propose our third model, RN-DMLP, which exhibits remarkable performance with sufficiently low MSE levels on S\&P 500 option data. We attribute the success of RN-DMLP to the increased flexibility of the distribution it can model. At last, we examine the risk-neutral moments and density graphs of RNDs, showing empirical findings that not only align with intuitions on the financial markets but also verify the rationality of the extracted RNDs. At last, through some analysis, we attribute the success of RN-DMLP to its ability to offer flexible term structures for risk-neutral skewness and kurtosis.

\bibliographystyle{rQUF}
\bibliography{GML_bibtex}

\appendices
\section{Proof of Proposition \ref{prop:GML-MLP}}
\label{sec:appendix1}

We require $\frac{\partial \hat{C}_{M}}{\partial \tau}\ge 0$ and $\frac{\partial \hat{P}_{M}}{\partial \tau}\ge 0$. The former is equivalent to
\begin{equation*}
	\begin{aligned}
		\frac{\partial \hat{C}_{M}}{\partial \tau} &= \frac{\partial}{\partial \tau} \left\{ S_te^{-r\tau} \frac{1}{N} \sum_{n=1}^{N} \left(e^{X(Z_n, \tau;\theta^M)} - S_t^{-1}K^C \right)^+ \right\} 
		\\ &= -rS_te^{-r\tau} \frac{1}{N} \sum_{n=1}^{N} \left(e^{X(Z_n, \tau;\theta^M)} - S_t^{-1}K^C \right)^+ + S_te^{-r\tau} \frac{1}{N} \sum_{n=1}^{N} \frac{\partial}{\partial \tau} \left(e^{X(Z_n, \tau;\theta^M)} - S_t^{-1}K^C \right)^+
		\\ &= S_te^{-r\tau} \frac{1}{N} \sum_{n=1}^{N} \frac{\partial}{\partial \tau} \left(e^{X(Z_n, \tau;\theta^M)} - S_t^{-1}K^C \right)\mathds{1}_{\{e^{X(Z_n, \tau;\theta^M)} \geq S_t^{-1}K^C\}} 
		\\ & \quad -rS_te^{-r\tau} \frac{1}{N} \sum_{n=1}^{N} \left(e^{X(Z_n, \tau;\theta^M)} - S_t^{-1}K^C \right)\mathds{1}_{\{e^{X(Z_n, \tau;\theta^M)} \geq S_t^{-1}K^C\}}
		\\ &= S_t e^{-r\tau} \frac{1}{N} \sum_{n=1}^{N} \left[\left(\frac{\partial X(Z_n, \tau;\theta^M)}{\partial \tau} - r\right) e^{X(Z_n, \tau;\theta^M)} + rS_t^{-1}K^C\right]\mathds{1}_{\{e^{X(Z_n, \tau;\theta^M)} \geq S_t^{-1}K^C\}} 
		\\ &= S_te^{-r\tau} \tilde{J}^{\tau}_{C}(\tau,K^C,\theta^M) \geq 0.
	\end{aligned}
\end{equation*}
Similarly, we have
\begin{equation*}
	\begin{aligned}
		\frac{\partial \hat{P}_{M}}{\partial \tau} &= \frac{\partial}{\partial \tau} \left\{ S_te^{-r\tau} \frac{1}{N} \sum_{n=1}^{N} \left(S_t^{-1}K^P - e^{X(Z_{n}, \tau;\theta^M)} \right)^+ \right\} 
		\\ &= -rS_te^{-r\tau} \frac{1}{N} \sum_{n=1}^{N} \left(S_t^{-1}K^P - e^{X(Z_{n}, \tau;\theta^M)} \right)^+ + S_te^{-r\tau} \frac{1}{N} \sum_{n=1}^{N} \frac{\partial}{\partial \tau} \left(S_t^{-1}K^P - e^{X(Z_{n}, \tau;\theta^M)} \right)^+
		\\ &= S_te^{-r\tau} \frac{1}{N} \sum_{n=1}^{N} \frac{\partial}{\partial \tau} \left(S_t^{-1}K^P - e^{X(Z_{n}, \tau;\theta^M)} \right)\mathds{1}_{\{e^{X(Z_{n}, \tau;\theta^M)} \leq S_t^{-1}K^P\}} 
		\\ & \quad -rS_te^{-r\tau} \frac{1}{N} \sum_{n=1}^{N} \left(S_t^{-1}K^P - e^{X(Z_{n}, \tau;\theta^M)} \right)\mathds{1}_{\{e^{X(Z_{n}, \tau;\theta^M)} \leq S_t^{-1}K^P\}}
		\\ &= S_t e^{-r\tau} \frac{1}{N} \sum_{n=1}^{N} \left[\left(r - \frac{\partial X(Z_{n}, \tau;\theta^M)}{\partial \tau}\right) e^{X(Z_{n}, \tau;\theta^M)} - rS_t^{-1}K^P\right]\mathds{1}_{\{e^{X(Z_{n}, \tau;\theta^M)} \leq S_t^{-1}K^P\}} 
		\\ &= S_te^{-r\tau} \tilde{J}^{\tau}_{P}(\tau,K^P,\theta^M) \geq 0.
	\end{aligned} 
\end{equation*}
In addition, the Constraint \ref{C6}, upon substituting the expecation by the sample average and $\mu(\tau)$ by $r\tau G^\mu(\tau;\theta^\mu)$ in Eqn. (\ref{eq:noarbitrage-general}),
\begin{equation}
	r\tau G^\mu(\tau;\theta^\mu) + \ln\left(\frac{1}{N} \sum_{n=1}^N e^{\sigma\sqrt{\tau} \cdot Z_n \cdot [G^Z(Z_n;\theta^Z) + G^\tau(\tau;\theta^\tau) + 1]}\right)  - r \tau = 0, \quad \forall \tau>0,\label{eq:constraint_mu_MLP}
\end{equation} 
is equivalent to setting $J^\mu_{M}(\tau,\theta^{M})=0$ where,
\begin{equation}\label{eqt:L2norm-MLP}
	J^\mu_{M}(\tau,\theta^{M}) = \left| r\tau \left( G^\mu(\tau;\theta^\mu) -1 \right)+ \ln\left(\frac{1}{N} \sum_{n=1}^N e^{\sigma\sqrt{\tau} \cdot Z_n \cdot [G^Z(Z_n;\theta^Z) + G^\tau(\tau;\theta^\tau) + 1]}\right) \right|^2.
\end{equation}

\section{Some Statistical Properties of Eqn. \eqref{eqt:ref_quantile}}
\label{appen:htqf}

In this section, we provide some statistical properties associated with the random variable defined in Eqn. \eqref{eqt:ref_quantile}, with a particular focus on its moment structure. Without loss of generality, we try to find the central moments of the random variable $X = Z(e^{uZ}/A + e^{-vZ}/A + 1)$, where $Z$ is a standard normal random variable and $u,v\ge 0$.

To find the expected value of the random variable $X$, we can use the linearity of expectation:
$$\mathbb{E}[X] = \mathbb{E}\left[ \frac{Z e^{uZ}}{A} + \frac{Z e^{-vZ}}{A} + Z \right] = \frac{1}{A}\mathbb{E}[Z e^{uZ}] + \frac{1}{A}\mathbb{E}[Z e^{-vZ}] + \mathbb{E}[Z].$$
As $\mathbb{E}[Z] = 0$, to evaluate the terms $\mathbb{E}[Z e^{cZ}]$ for any constant $c$, we can use the Moment Generating Function (MGF) of $Z$, which is $M_Z(t) = \mathbb{E}[e^{tZ}] = e^{t^2/2}$.
By differentiating both sides with respect to $t$, we can bring down a $Z$:
$$\frac{d}{dt} \mathbb{E}[e^{tZ}] = \frac{d}{dt} \left( e^{t^2/2} \right),\quad\text{equivalently,}\quad\mathbb{E}[Z e^{tZ}] = t e^{t^2/2}.$$
Now, applying this result to our specific terms where $t = u$ and $t = -v$,
and substituting these expectations back into our original equation:
$$\mathbb{E}[X] = \frac{1}{A} \left( u e^{u^2/2} \right) + \frac{1}{A} \left( -v e^{v^2/2} \right) + 0 = \frac{u e^{u^2/2} - v e^{v^2/2}}{A}.$$

To find the variance of the random variable $X$, we use the standard variance formula: $\text{Var}(X) = \mathbb{E}[X^2] - (\mathbb{E}[X])^2$.
As $\mathbb{E}[X] = \frac{u e^{u^2/2} - v e^{v^2/2}}{A}$, now, we need to calculate $\mathbb{E}[X^2]$.
We first expand $X^2$:
$$X^2 = Z^2 \left( \frac{e^{2uZ}}{A^2} + \frac{e^{-2vZ}}{A^2} + 1 + \frac{2e^{(u-v)Z}}{A^2} + \frac{2e^{uZ}}{A} + \frac{2e^{-vZ}}{A} \right).$$
Using the linearity of expectation, $\mathbb{E}[X^2]$ requires us to evaluate terms of the form $\mathbb{E}[Z^2 e^{cZ}]$ for various constants $c$.
Just like we used the first derivative of the MGF to find $\mathbb{E}[Z e^{cZ}]$, we can use the second derivative of the MGF to find $\mathbb{E}[Z^2 e^{cZ}]$:
$$\frac{d^2}{dt^2} \mathbb{E}[e^{tZ}] = \frac{d}{dt} \left( t e^{t^2/2} \right),
\quad\text{equivalently,}\quad\mathbb{E}[Z^2 e^{tZ}] = (1 + t^2) e^{t^2/2}.$$
Now we can apply this formula to each part of our expanded $X^2$ expression and construct the final variance:
\begin{align*}
	\text{Var}(X) = &  \frac{(1 + 4u^2) e^{2u^2} + (1 + 4v^2) e^{2v^2} + 2(1 + (u-v)^2) e^{(u-v)^2/2}}{A^2} + \frac{2(1 + u^2) e^{u^2/2} + 2(1 + v^2) e^{v^2/2}}{A} \\
	& + 1  -  \frac{u^2 e^{u^2} - 2uv e^{(u^2+v^2)/2} + v^2 e^{v^2}}{A^2} .
\end{align*}

To find the exact expressions for the skewness and kurtosis of $X$, we would need to calculate the third and fourth central moments. Since $X = Z(e^{uZ}/A + e^{-vZ}/A + 1)$ is a trinomial, expanding $X^3$ and $X^4$ results in a massive number of terms (10 terms for the cubic expansion and 15 for the quartic), making the raw algebra quite unwieldy. Just as we used the first and second derivatives of the MGF $M_Z(t) = e^{t^2/2}$ to find the mean and variance, we can use the third and fourth derivatives to evaluate the terms in $X^3$ and $X^4$. Let's skip these tedious but straightforward details. Instead, our analysis centers on deriving mathematical insights concerning the structural properties of the distribution of $X$.

Skewness measures the asymmetry of the distribution. The relationship between $u$ and $v$ entirely dictates the skewness of $X$. When $u = v$, the skewness is exactly 0.
If $u > v$, the term $e^{uZ}$ grows much faster for positive values of $Z$ than $e^{-vZ}$ does for negative values of $Z$. This stretches the right side of the distribution significantly.  The distribution will have a positive (right) skew.
If $u < v$, the opposite happens. The $e^{-vZ}$ term amplifies the negative values of $Z$, creating a longer tail on the left. The distribution will have a negative (left) skew.

Kurtosis measures the tailedness of a distribution. A standard normal distribution has a kurtosis of 3 (or an excess kurtosis of 0). 
Look at the behavior of $X$ as $Z$ gets very large. For a large positive $Z$, $X$ behaves roughly like $\frac{Z e^{uZ}}{A}$. Because of the exponential term, $X$ will generate extreme outliers far more frequently than a standard normal distribution would. 
For any $u > 0$ or $v > 0$, the distribution of $X$ will have heavy tails and a sharply peaked center compared to a Gaussian curve. This means the kurtosis will be strictly greater than 3 (positive excess kurtosis). As the parameter $u$ increases, the right tail of the distribution exhibits greater heaviness; an analogous relationship holds for parameter $v$ with respect to the left tail.


\section{Deep Models as Benchmarks}
\label{appen:deep}

\begin{figure}[t]
	\centering 
	\includegraphics[width=.85\linewidth]{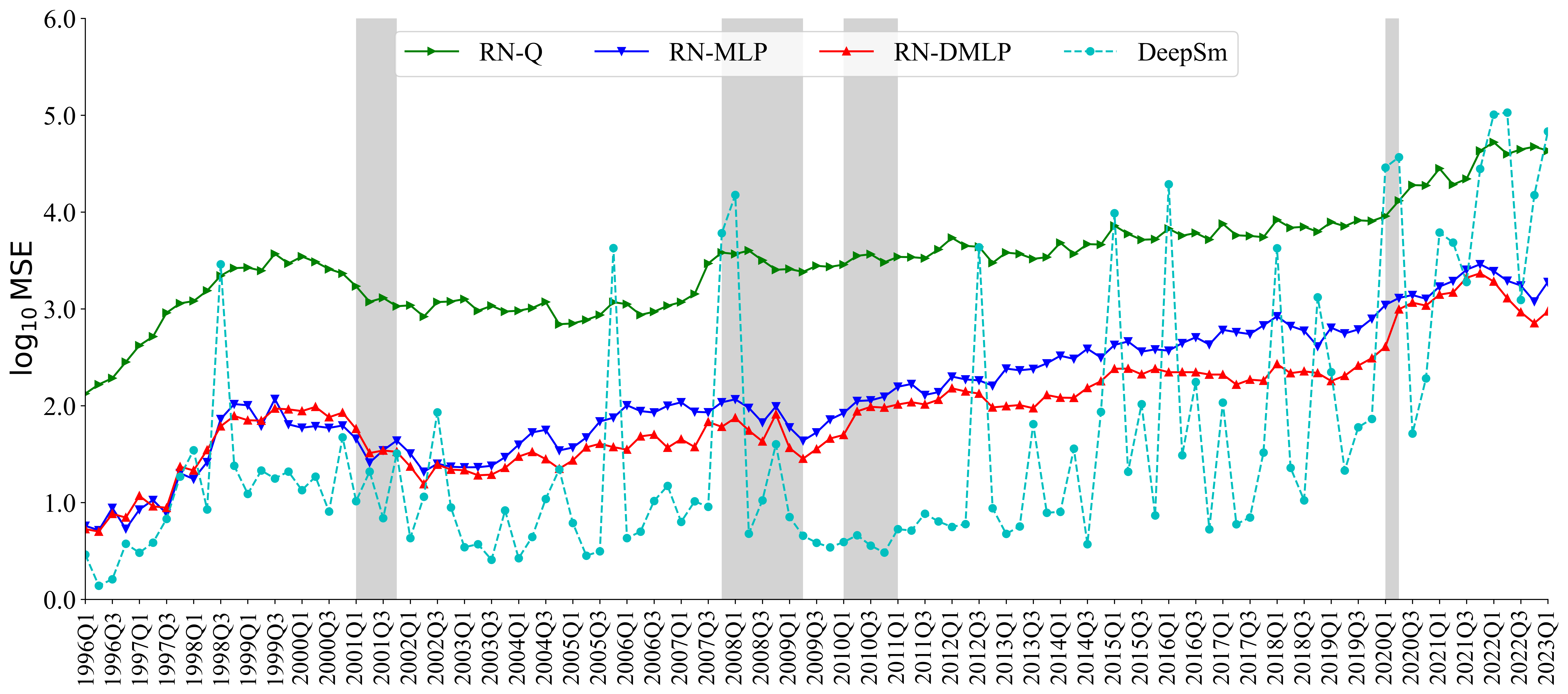}
	\caption{Pricing performance under the multiple $\tau$ setting. The $\log_{10}\text{MSE}$ (averaged in every quarter) on the testing set is plotted against varying time for each model. The shaded areas represent the financial crisis periods. Overall, RN-DMLP achieves the best performance and DeepSm yields unstable performance.}
	\label{fig:baseline_deepsm}
\end{figure}

\begin{table}[t]
	\caption{Comparison of pricing performance averaged over all years under the multiple $\tau$ setting for our models and DeepSm.}
	\label{tab:baseline_deepsm}
		\centering
		\begin{tabular}{llcccc}
			\toprule
			{} & {} & RN-Q & RN-MLP & RN-DMLP & DeepSm \\
			\midrule
			\multirow{2}{*}{Testing} & {MSE} & 9263.23 & 546.69 & \textbf{348.64} & 6927.91 \\
			{} & {Relative} & 4472.28 & 3.49 & \textbf{0.43} & 2.77 \\
			\multirow{2}{*}{Extreme} & {MSE} & 1742.48 & 562.49 & \textbf{391.10} & 7166.23 \\
			{} & {Relative} & 2277.36 & 4.98 & \textbf{1.11} & 47.37 \\
			\bottomrule
		\end{tabular}
\end{table}

Complementing benchmarks with additional classes of models—such as neural-network-based models or more flexible IVS specifications—would further strengthen the empirical studies. In this section,
we add an additional neural-network-based volatility-surface model to our benchmark suite: \textit{Deep Sm}oothing of the Implied Volatility Surface, as introduced in \cite{ackerer2020deep}. This paper is one of the earlier but highly representative neural-network-based approaches to option pricing and implied volatility surface modeling.

\paragraph{(1) Experimental setting for the new benchmark DeepSm}
We evaluate the DeepSm model under exactly the same multiple-maturity (multiple-$\tau$) setting as our RN-DMLP architecture, using both the \textit{Testing Set} and the \textit{Extreme-Moneyness Set}. The empirical results are summarized in Table \ref{tab:baseline_deepsm} and Figure \ref{fig:baseline_deepsm}.

\paragraph{(2) Analysis of the results}
Across both sets, the RN-DMLP model achieves substantially lower average pricing errors compared with DeepSm. On the \textit{Testing Set}, RN-DMLP attains an MSE of \textbf{348.64}, which is significantly lower than the \textbf{6927.91} obtained by DeepSm.  On the \textit{Extreme-Moneyness Set}, the discrepancy is even more pronounced: \textbf{391.10} for RN-DMLP versus \textbf{7166.23} for DeepSm. From Figure \ref{fig:baseline_deepsm}, we can see that the DeepSm model exhibits higher fluctuation compared with RN-DMLP. While DeepSm occasionally achieves short intervals of relatively low error, its pricing performance is highly unstable over time, exhibiting large spikes and strong temporal fluctuations. In contrast, RN-DMLP produces consistently low and stable loss trajectories, indicating a more reliable generalization capability.

This pattern aligns well with the conceptual distinction between the two approaches. DeepSm primarily focuses on smoothing the implied volatility surface, whereas RN-DMLP directly models the risk-neutral densities across different time-to-maturities using a distributionally structured architecture, which appears to be more robust to variations in market conditions.

\section{Violation of Conditions and Model Modification}
\label{appen:violation}

We conducted two sets of additional experiments to (1) quantify how strongly the condition $J^{\mu} = 0$ is violated in practice, and (2) investigate whether alternative drift specifications—specifically a flexible, learnable drift or a hard constraint enforcing—can improve calibration.

\begin{figure}[t]
	\centering 
	\subfigure[Histogram of $\log_{10} \left| J^{\mu} \right| $.]{\includegraphics[width=.45\linewidth]{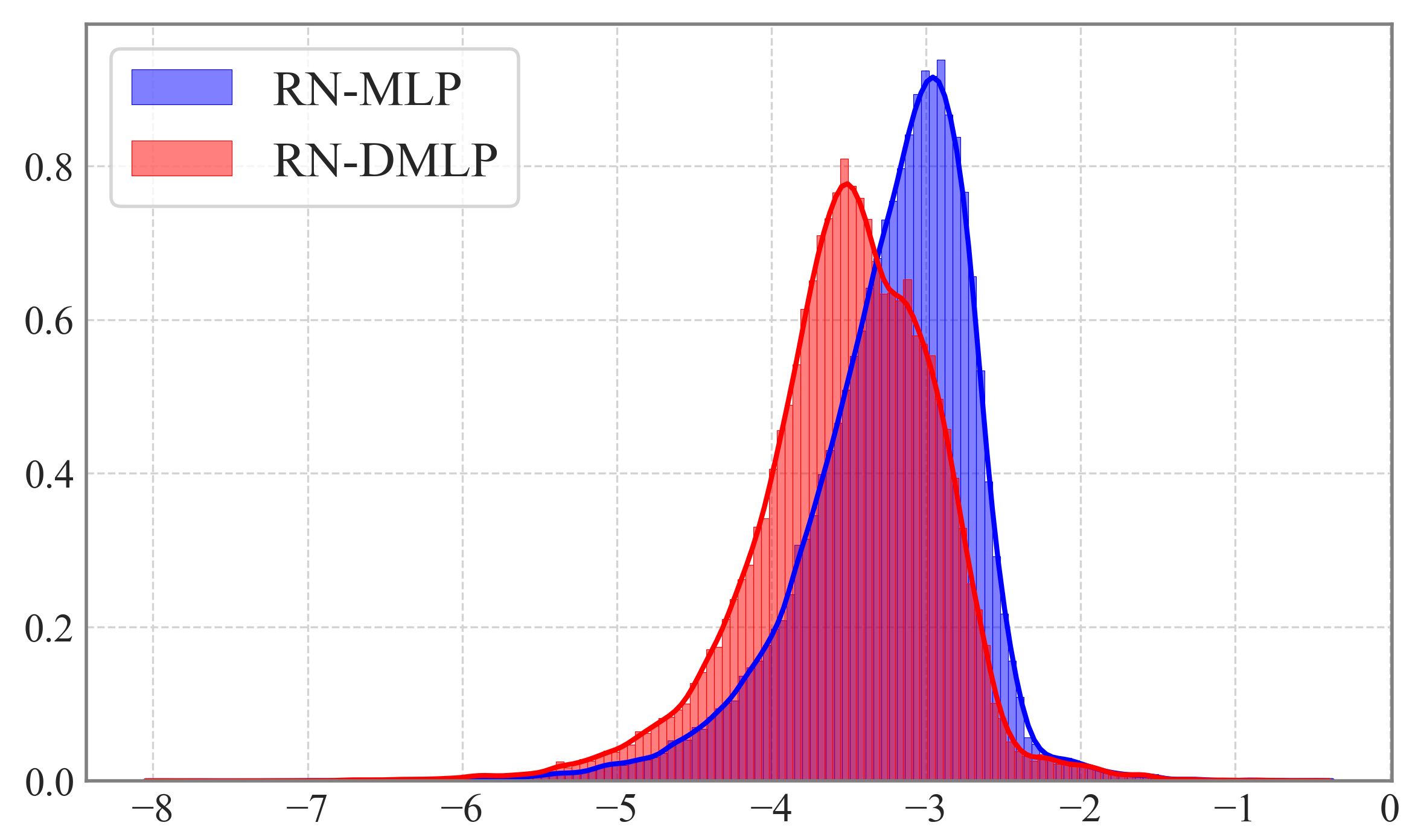}}
	\subfigure[Scatter plot of $J^{\mu}$ versus the out-of-sample MSE.]{\includegraphics[width=.45\linewidth]{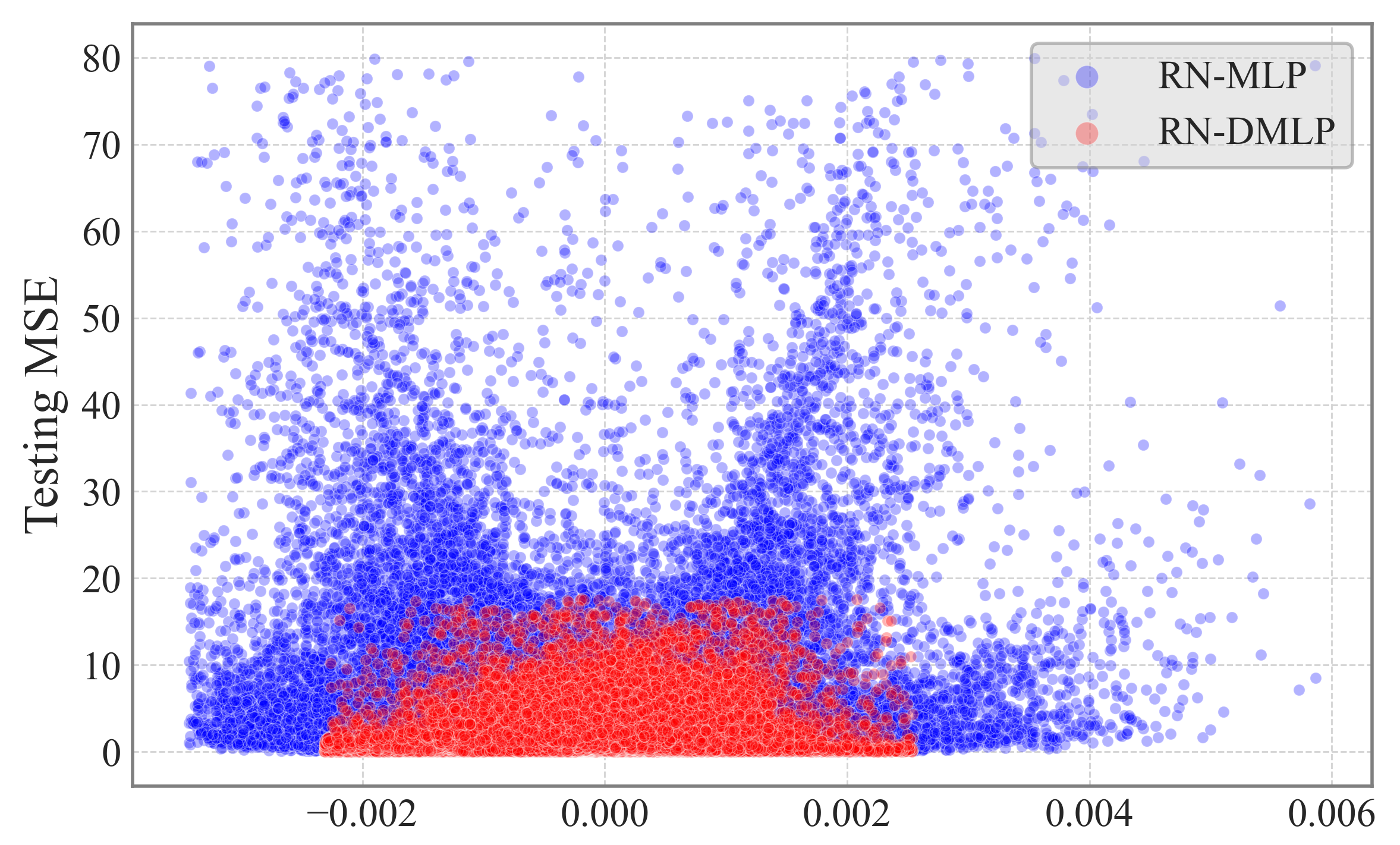}}
	\caption{Assessment of the violation of the constraint $J^{\mu}=0$. Subfigure (a) displays the histogram of $\log_{10}|J^{\mu}|$, demonstrating that its magnitudes are generally close to zero and the condition $J^{\mu}=0$ is well satisfied in practice. Subfigure (b) illustrates the relationship between $J^{\mu}$ and the out-of-sample testing MSE, showing no evident correlation. Generally, RN-DMLP exhibits lower MSE values and better satisfaction of $J^{\mu}=0$ compared with RN-MLP.
	}
	\label{fig:Jmu_desc}
\end{figure}

\paragraph{(1) Assessment of the violation of the constraint $J^{\mu} = 0$}
We first examined the values of $|J^{\mu}|$ generated by the already-trained RN-MLP and RN-DMLP models. For each observation date and each time-to-maturity $\tau$, a corresponding value of $|J^{\mu}|$ exists.
In Figure \ref{fig:Jmu_desc}, Subfigure (a) presents the histogram of $\log_{10} \left| J^{\mu} \right| $, and Subfigure (b) shows the scatter plot of $J^{\mu}$ against the corresponding out-of-sample MSE.

The results consistently confirm that violations of the condition $J^\mu = 0$ are slight in practice for both models. The visual evidence in Figure \ref{fig:Jmu_desc}(a) demonstrates this high level of satisfaction, with the distribution of $\log_{10} \left| J^{\mu} \right| $ being heavily concentrated around $-3$ or $-4$. Critically, the histogram clearly shows that the RN-DMLP's distribution is shifted further to the left, indicating tighter clustering and slighter constraint violations compared to RN-MLP. This qualitative observation is strongly supported by the summary statistics presented in Table \ref{tab:Jmu_summary}: the RN-DMLP model exhibits a negligible median violation of only $0.022 \times 10^{-3}$. Furthermore, the RN-DMLP model demonstrates a substantially tighter concentration of $J^{\mu}$ values compared to RN-MLP. Therefore, the trained models remain very close to the theoretically correct values.

\begin{table}[t]
	\caption{Summary statistics of $J^{\mu}$ ($\times 10^{-3}$) for RN-MLP and RN-DMLP models.}
	\label{tab:Jmu_summary}
	\centering
	\begin{tabular}{lccccccc}
		\toprule
		& Mean & Std & $Q_{1\%}$ & $Q_{25\%}$ & Median & $Q_{75\%}$ & $Q_{99\%}$ \\
		\midrule
		RN-MLP & -0.053 & 4.800 & -3.458 & -0.965 & -0.169 & 0.468 & 6.152 \\
		RN-DMLP & 0.060 & 4.850 & -2.571 & -0.450 & -0.022 & 0.252 & 5.832 \\
		\bottomrule
	\end{tabular}
\end{table}

\paragraph{(2) Does using a flexible or hard-constrained drift help calibration?}
To directly answer this question, we implemented two variants of our model:

(i) Flexible drift variant:
\begin{equation}
	\begin{aligned} 
		X_\tau^{F} & :=X(Z, \tau;\theta^{F}) = \mu^{F} \tau G^\mu(\tau;\theta^\mu) \\
		& + \sigma\sqrt{\tau} Z [G^Z(Z;\theta^Z) + G^\tau(\tau;\theta^\tau) + 1], \quad \forall \tau>0, \\ 
	\end{aligned}
\end{equation}
where the trainable parameter set is $\theta^{F} = \{\mu^F, \theta^{\mu}, \sigma, \theta^Z, \theta^\tau\}$. We refer this variant to RN-MLP-F. The RN-DMLP-F model is defined analogously as a combination of two RN-MLP-Fs.

(ii) Hard constraint variant:
\begin{equation}
	\begin{aligned} 
		X_\tau^{H} & :=X(Z, \tau;\theta^{H}) = \mu^{H} + Q( Z,\tau; \theta^{\sigma}) \\
		& = \mu^{H} + \sigma\sqrt{\tau} Z [G^Z(Z;\theta^Z) + G^\tau(\tau;\theta^\tau) + 1], \quad \forall \tau>0, \\ 
	\end{aligned}
\end{equation}
where $\mu^{H} = r\tau - \ln\left(\frac{1}{N}\sum^N_{n=1} e^{ Q( Z_n,\tau; \theta^{\sigma})} \right)$. We refer this variant to RN-MLP-H. The RN-DMLP-H model is defined analogously as a combination of two RN-MLP-Hs.

Table \ref{tab:Jmu_model_error} presents the pricing MSE for our model and the variants. The original RN-DMLP model remains the best-performing architecture across both tests. The flexible drift variant (-F) worsens the pricing accuracy (e.g., RN-MLP-F's $\text{MSE}$ jumps from $3.052$ to $12.191$). This suggests that allowing $\mu$ to be freely learned introduces unnecessary complexity and degrees of freedom that destabilize the training process. Although the hard constraint variant (-H) is theoretically sound, it leads to inferior calibration performance (RN-DMLP-H's $\text{MSE}$ is $4.941$ versus $1.292$ for the original RN-DMLP). Therefore, applying strict theoretical constraints to deep learning architectures is a critical challenge.

Table \ref{tab:Jmu_model_summary} presents the summary statistics of $J^{\mu}$ for our model and the variants. As expected, the -H variants satisfy $J^{\mu}=0$ exactly. Moreover, the -F variants exhibit large dispersion in $J^{\mu}$, indicating that the flexible drift term does not help to satisfy the constraint. The original RN-MLP and RN-DMLP models already obtain sufficiently small $J^{\mu}$ values. 

\begin{table}[t]
	\caption{Pricing MSE for RN-MLP, RN-DMLP, and their variants -F and -H.}
	\label{tab:Jmu_model_error}
	\centering
	\begin{tabular}{lcccccc}
		\toprule
		{} & RN-MLP & RN-MLP-F & RN-MLP-H & RN-DMLP & RN-DMLP-F & RN-DMLP-H \\
		\midrule
		Testing & 3.052 & 12.191 & 3.943 & 1.292 & 2.859 & 4.941 \\
		Extreme & 16.669 & 26.012 & 13.304 & 11.322 & 14.102 & 13.228 \\
		\bottomrule
	\end{tabular}
\end{table}

\begin{table}[t]
	\caption{Summary statistics of $J^{\mu}$ ($\times 10^{-3}$) for RN-MLP, RN-DMLP, and their variants -F and -H.}
	\label{tab:Jmu_model_summary}
	\centering
	\begin{tabular}{lccccc}
		\toprule
		Model & Mean  & Median  & Std  & $Q_{25\%}$  & $Q_{75\%}$  \\
		\midrule
		RN-MLP & -0.184 & -0.146 & 0.377 & -0.474 & 0.121 \\
		RN-MLP-F & 0.069 & 0.203 & 1.540 & -0.942 & 1.036 \\
		RN-MLP-H & 0.000 & 0.000 & 0.000 & 0.000 & 0.000 \\
		RN-DMLP & -0.045 & -0.020 & 0.186 & -0.185 & 0.109 \\
		RN-DMLP-F & 0.014 & 0.029 & 0.384 & -0.183 & 0.192 \\
		RN-DMLP-H & 0.000 & 0.000 & 0.000 & 0.000 & 0.000 \\
		\bottomrule
	\end{tabular}
\end{table}

\end{document}